\documentstyle [11pt,epsf]{article}

\begin{document}
\newcommand{\EqRef}[1]{(\ref{eqn:#1})}
\newcommand{\Abs}[1]{|#1|}
\newcommand{\expct}    [1]{\left\langle {#1} \right\rangle}
\newcommand{\Tr}[0]{\mbox{Tr}}

\title{Error of semiclassical eigenvalues in the semiclassical limit - \\
An asymptotic analysis of the Sinai billiard}
\author{Per Dahlqvist \\
Mechanics Department \\
Royal Institute of Technology, S-100 44 Stockholm, Sweden\\[0.5cm]
}
\date{}
\maketitle

\begin{abstract}
We estimate the error in the semiclassical trace formula for the Sinai
billiard under the assumption that the largest source of error is due to
Penumbra diffraction, that is diffraction effects for trajectories
passing within a distance $R \cdot O((kR)^{-2/3})$ to the disk and trajectories
being scattered in very forward directions. Here $k$ is the momentum and
$R$ the radius of the scatterer.
The semiclassical error is estimated by perturbing the Berry-Keating formula.
The analysis necessitates an asymptotic analysis of very long periodic
orbits. This is obtained within an approximation originally due to
Baladi Eckmann and Ruelle.
We find that the average error, for sufficiently large value of $kR$, will exceed
the mean level spacing.
\end{abstract}

\section{Introduction}
\label{s:intro}

During the early days of the trace formula nobody really believed that
it could be used to predict individual eigenvalues,
at least not in the strict semiclassical limit. 
There are mainly two (related) sources of errors.
First, the semiclassical energy domain Green function is obtained by
Laplace transforming the Van-Vleck propagator \cite{Gut}.
However, quantum evolution follows classical evolution only for a limited time,
a time that seems to be longer than first expected \cite{Heller}, but still
limited. Computing the Laplace transform (with time going from zero
to infinity) of such an object is of course adventurous.

Secondly, the trace formula is obtained by taking the trace of this energy domain
Green function by stationary phase technique. 
This is the procedure that selects out the periodic
orbits. 
Whether or not this stationary phase approximation is justified for a particular
cycle depends on $\hbar$, for sufficiently small $\hbar$ 
it is always justified. So, the
the performance of the trace formula depends
on the set of cycles that are required 
to resolve a particular state, and the accuracy of their
semiclassical weights.

Consequently, the {\em semiclassical error} depends on the 
context, the method by which the eigenvalues are extracted. 
In the Berry-Keating
method\cite{BK90} it depends only on the error of the amplitudes of
periodic orbits up to certain length,
whereas in complex methods, based on 
{\em cycle expansions} \cite{CE93},
the error of the long cycles effects the convergence of the cycle expansion
and the final position of the eigenvalue.
An alternative approach is based on the boundary integral method and
the semiclassical limit of the characteristic determinant\cite{Boas}.
Bogomolny's transfer matrix method is closely related to this approach\cite{Bogo92}.

However, it turned out 
from, numerical computations
that the Gutzwiller-Voros zeta function does exhibit
complex zeroes quite close to the (real) quantum eigenvalues, at least in the lower part
of the spectrum\cite{DahlRuss91,Freddy92,Tanner91}.
The Berrry-Keating formula performed even better\cite{Tanner91,SS91}.
This approach is based on a functional equation for the (exact) spectral
determinant.
By insisting on using the functional equation in the semiclassical limit,
one actually put in information into the computation: {\em the spectrum is real}.
The result may be a quite respective 
number of eigenvalues, only a few percent wrong.

None of these computations indicated what will happen in the strict semiclassical limit.
Common estimates have suggested that the semiclassical errors,
measured relative the mean level spacing, should tend to a constant
as $\hbar \rightarrow \infty$, 
for system in two degrees of freedom. For a nice review, see ref \cite{Prim98}.

Two common features of 
chaotic systems may cause problems to the stationary phase approximation.
The first is intermittency. From a periodic pont of view, it means that
long orbits does not need to be very unstable,
or more precisely, periodic orbit stabilities cannot be exponentially bounded
with their length. If a long cycle is "relatively stable" it means that 
the corresponding saddle point 
is not isolated enough from its surroundings.

Secondly, the complex topology of generic systems make bifurcations abundant
(with respect to variation of some parameter). 
So, for a generic systems there are "almost
existing" and "almost forbidden" orbits all over phase space. For billiards
the trace integral scans a function sprinkled with discontinuities.
Nearly pruned orbits live near such discontinuities and the saddle point integrals has
to be provided with cutoffs which leads to diffraction effects.
Even Axiom-A systems such as the Baker map \cite{Saraceno} suffers from
diffraction effects but to a less extent.
For smooth potentials one has to face stabilization of cycles close to bifurcation.
Sufficiently pruned orbits can be included as ghost orbits \cite{ghost} 
but close to bifurcations
uniform approximations has to be invoked\cite{Sieb97,Sieb97a}.\marginpar{}

There are two pioneering studies indicating that many of the terms included in the 
Berry-Keating sum are way off, 
the stationary phase approximation behind them
is simply not justified.
Tanner \cite{Tanner} studied the 
dynamics close to the bouncing ball orbit in the
stadium billiard, and Primack et.al. \cite{Prim95} studied 
the Sinai billiard and
orbits scattering in very
forward direction or sneaking very close to the disk. Both cases deals with
systems with neutral orbits and as such intermittent.

The paper \cite{Prim95} by Primack et.al. is the main inspiration for the present study.
By estimating the size of this {\em penumbra} and its scaling in energy, 
they concluded that {\em "the semiclassical approximation fails for the
majority of the relevant PO's} (periodic orbits) {\em in the semiclassical limit".}
It thus seems likely that
the Berry-Keating approach will eventually cease to produce individual eigenvalues.
Such a conclusion is by no means obvious. 
Simple conclusions can (maybe) be drawn if the system is uniformly hyperbolic.
But cycles contribute with very different weights in intermittent systems,
such as the Sinai billiard.
If a minority of non diffractive cycles carried a large part of the semiclassical
weight one could perhaps argue that 
the trace formula could be saved. But unfortunately the situations is the opposite.
The cycles being most prone to penumbra diffraction have large semiclassical weights.
The goal of this paper is to make this reasoning more precise.

In a latter study Primack et.al. \cite{Prim98} seems to tone down the importance of 
Penumbra diffraction. They suggest that the semiclassical error
in the semiclassical limit is of the order of the mean level spacing,
or at most diverges logarithmically. It could thus be possible to resolve
individual states even in the strict semiclassical limit.
The study is mainly numerical, it is an ambiguous attempt to extrapolate
from finite sets of quantum states and periodic orbits into the semiclassical limit. 
Our present study will not support their claims. We will indeed find that
the error will irrevocably increase beyond the mean level spacing.

It is evident that a numerical study of the semiclassical error with
periodic orbits is, least to say, difficult.
A central tool in our approach will be 
asymptotic theory for the sets of periodic
orbits based on an idea of Baladi Eckmann and Ruelle \cite{BER,PDreson,PDsin,PDzak,PDlyap}.

What we do is a model study of an intermittent system
in two degrees of freedom. 
One should of course be cautious when trying to 
generalize the result to other system,
especially to hyperbolic ones.
The trace formula is indeed exact for some hyperbolic system,
such as the Cat map \cite{CAT1,CAT2} 
and geodesic flow on surfaces of constant
negative curvature \cite{geo}. 
The trace formula will probably perform
much better in the hyperbolic case than in the intermittent. 

The disposition of the article is very much like a cooking recipe.
In section \ref{s:ingred} we present all the ingredients, such as
the semiclassical zeta function, the Berry-Keating formula, penumbra
diffraction, some classical periodic orbit theory and the BER approximation.
In sec.\ \ref{s:prep} we do the actual cooking. We consider the shift of a
zero of the Berry-Keating formula if a perturbation, due to an error, is added.
This simple exercise gives us the semiclassical error
in terms of finite sums over pseudo
orbits. 
We then relate these pseudo orbit sums
to various zeta functions. 
These zeta functions
are then calculated in the asymptotic theory which we call the
BER approximation.
In section \ref{s:result} we present the outcome of these exercises. 
Then follows (section \ref{s:disc}) a round table discussion 
about the validity of the various assumptions and approximations that 
underlies the results.

\section{Ingredients}
\label{s:ingred}
\subsection{The semiclassical zeta function}
\label{s:GV}

The starting point will be the Gutzwiller-Voros zeta function \cite{Gut}
whose zeros is to be associated with the quantum eigenvalues.
It is represented as a product over all primitive periodic orbits $p$
of the systems.
\begin{equation}
Z^{sc}(E)=\prod_{p}\prod_{m=0}^{\infty}
          \left(1-\frac{e^{i\left[S_{p}/\hbar-\mu_{p}\frac{\pi}{2}\right]}}
            {|{\Lambda_{p}}|^{1/2}\Lambda_{p}^{m}}\right)
          \ \ .    \label{eqn:ZGV}
\end{equation}
where $S_p$ is the action along $p$,
$\mu_p$ the Maslov index and
$\Lambda_p$ is the expanding eigenvalue of the Jacobian.
To turn this into a Dirichlet series one first expand the inner Euler-
product
\begin{equation}
       Z^{sc}(E)=\prod_{p}\sum_{n=0}^{\infty}
            \frac{\Lambda_{p}^{-n\left(n-1\right)/2}}
                 {\prod_{j=1}^{n}\left(1-\Lambda_{p}^{-j}\right)}
            \left( - \frac{e^{i\left[S_{p}/\hbar-\mu_{p}\frac{\pi}{2}\right]}}
            {|{\Lambda_{p}}|^{1/2}} \right)^{n} \ \ 
\end{equation}
\begin{equation}
       =\prod_{p}\sum_{n=0}^{\infty}C_{p,n}
       e^{i\left[nS_{p}
       -n\mu_{p}\frac{\pi}{2}\right]} \ \ .
\end{equation}
where
\begin{eqnarray}
C_{p,n}=(-1)^n \frac{\Lambda_{p}^{-n\left(n-1\right)/2}}
  {|{\Lambda_{p}}|^{n/2}\prod_{j=1}^{n}\left(1-\Lambda_{p}^{-j}\right)}
\end{eqnarray}

If we now expand the product over $p$, 
we obtain a {\em cycle expansion} \cite{DasBuch} -
a sum
over all pseudo-orbits, that is all
distinct combinations of periodic orbits:
$\alpha=p_1^{n_{p_1}} p_2^{n_{p_2}} \ldots p_k^{n_{p_k}}\ldots$

\begin{equation}
      Z^{sc}(E)=\sum_{\alpha}C_{\alpha}e^{i
\left[S_{\alpha}-\mu_{\alpha}\frac{\pi}{2}\right]}
       \ \ , \label{eqn:GVcycexp}
\
\end{equation}
where we have defined the quantities
\begin{equation}
        C_{\alpha}=\prod_{p}C_{p,n_{p}}  \ \ ,  \label{eqn:Calpha}
\end{equation}
\begin{equation}
        S_{\alpha}=\sum_{p}n_{p}S_{p}  \ \ ,   \label{eqn:Salpha}
\end{equation}
\begin{equation}
        \mu_{\alpha}=\sum_{p}n_{p}\mu_{p}  \ \ .   \label{eqn:mualpha}
\end{equation}

We will restrict ourselves to billiards, the cycle action 
$S_p$ is then given in terms of the geometric 
length $S_p=L_p \cdot k$ where $k=\sqrt{2E}$.
The units are chosen
such that $m=\hbar=1$ and the semiclassical limit $\hbar \rightarrow 0$ is replaced by
$k \rightarrow \infty$.
In the following we absorb the maslov indices in the amplitudes $C_\alpha$
\begin{equation}
Z^{sc}(k) = \sum_\alpha C_\alpha e^{iL_\alpha k}
\end{equation}
Since we are considering billiards, the redefined amplitudes $C_\alpha$ will still be real.

Note that the size of the amplitudes is
\begin{equation}
C_{p,n}\sim \frac{1}{|\Lambda|^{n^2/2}}
\end{equation}
to leading order, and thus decays fast with $n$.
The zeta function is not seriously affected if one restricts the
$n$'s to $n \in \{ 0,1\}$. This amounts to retain only the factor
$m=0$ in \EqRef{ZGV}. The resulting type of zeta function is often referred
to as a {\em dynamical zeta function}.

\subsection{The Berry-Keating formula}

\label{s:BK}

The spectral determinant for a billiard obeys the
functional equation
\begin{equation}
D(k)=D(-k)  \  \  . \label{eqn:funceq}
\end{equation}
The semiclassical analogue to the spectral determinant is
\begin{equation}
D^{sc}(k)= e^{-i\pi \bar{N}(k)}Z^{sc}(k)=
\sum_\alpha C_\alpha e^{i(L_\alpha k-\pi \bar{N}(k) )} \label{eqn:specdet}
\end{equation}
The idea of Berry and Keating \cite{BK90} was to
postulate that this semiclassical determinant also satisfies the functional
equation 
\EqRef{funceq}. This is of course not exactly true 
\footnote{In e.g. the Sinai billiard the semiclassical
zeta function has a branch cut
along the negative imaginary k-axis and the equation
$D^{sc}(k)=D^{sc}(-k)$ cannot hold \cite{PDnaka}}
but by insisting on it
one can convert eq. \EqRef{specdet} to a finite sum
\begin{equation}
D^{sc}(k)=2\sum_{\alpha : L_\alpha<L_{BK}}C_\alpha \cos (L_\alpha k-\pi \bar{N}(k) )
\label{eqn:BK}
\end{equation}
where
\begin{equation}
L_{BK}=\pi \frac{d\bar{N}(k)}{dk}  \   \   .
\label{eqn:LBK}
\end{equation}
$\bar{N}(k)$ is the mean spectral staircase function. For a billiard it is,
to leading order, given by
\begin{equation}
\bar{N}=\frac{A k^2}{4\pi}  \  \  ,
\end{equation}
where $A$ is the billiard area.
So the cutoff length is given by
\begin{equation}
L_{BK}=\frac{Ak}{2} \label{eqn:LBKbill}
\end{equation}

If neutral orbits are present, and they are in the Sinai billiard, their
contribution can be included in $\bar{N}(k)$ which is then decorated by
oscillation whose amplitude decreases with increasing $k$.

\subsection{The classical zeta function}
\label{s:traces}

Another central object in our investigation will be
the (weighted) 
evolution operator\cite{DasBuch}, whose action on a phase
space distribution function $\Phi(x)$ is given by
\begin{equation}
{\cal L}_w^t  \Phi(x)=\int w(x,t)\delta (x-f^t(y))\Phi(y)dy   \ \ .
\label{eqn:evol}
\end{equation}
The phase space point $x$ is taken by the flow to $f^t(x)$ during time
$t$. $w(x,t)$ is a weight associated with a trajectory starting at $x$  
and evolved during time $t$. It is multiplicative along the flow, that
is $w(x,t_1+t_2)=w(x,t_1)w(f^{t_1}(x),t_2)$.
If $w\equiv 1$, the operator just describes the classical evolution of the 
phase space density.

In the following we will restrict ourselves to chaotic 2-D billiards,
and will use traversed length $L$ as 'time' variable.

The trace of this operator can be represented in terms of periodic orbits
in two ways.
First as a sum
\begin{equation}
tr {\cal L}_w^L  =
\sum_p L_p \sum_{r=1}^{\infty} w_p^r \frac{\delta(L-rL_p)}
{\Abs{det(1-M_p^r)}}  \ \ ,
\label{eqn:tracedelta}
\end{equation}
where $r$ is the number of repetitions of primitive orbit $p$, having period 
$L_{p}$, and  $M_{p}$ is the Jacobian of the Poincar\'{e} map, 
its expanding eigenvalue is $\Lambda_p$ and $w_p$ is the weight
associated with cycle $p$.

The trace can also be written in terms of
of a {\em zeta function}
or {\em Fredholm determinant}:
\begin{equation}
tr {\cal L}_w^L = \frac{1}{2\pi i}
\int_{a-i\infty}^{a+i\infty} e^{sL}\frac{Z_w'(s)}{Z_w(s)}ds   \ \ .
\label{eqn:traceZ}
\end{equation}
The classical zeta function $Z_w(s)$ is given by
\begin{equation}
     Z_w(s)=\prod_{p}\prod_{m=0}^{\infty}
          \left(1-w_p \frac{e^{-sL_{p}}}
    {\Abs{\Lambda_{p}} \Lambda_{p}^{m}}\right)^{m+1}
          \ \ .                      
\end{equation}

Again we we suffice with the $m=0$ factors, and  define instead the classical
zeta function as
\begin{equation}
     Z_w(s)=\prod_{p}
          \left(1-w_p \frac{e^{-sL_{p}}}
    {\Abs{\Lambda_{p}}}\right)
          \ \ .                       \label{eqn:Zw}
\end{equation}

In case the zeta function is entire, the trace can be written as a sum over
zeros $s_\gamma$ of the zeta function
\begin{equation}
tr {\cal L}_w^L = \sum_\gamma e^{s_\gamma L} \ \  ,  \label{eqn:trsgamma}
\end{equation}
where $e^{s_\gamma L} $ can be interpreted as the eigenvalues of the
evolution operator.
Precisely as  the semiclassical zeta function, the classical zeta function
$Z_w(s)$ can be subject to a cycle expansion
\begin{equation}
Z_w(s)= \sum_\alpha a_\alpha(w) e^{-s L_\alpha}  \  \    .
\label{eqn:clcycleexp}
\end{equation}
We will eventually use the weight $w(x,L)$ to account for diffraction 
but for the time being it is just an arbitrary weight.

\subsection{The BER approximation}
\label{s:BERtheory}

We note that the identity in sec \ref{s:traces}, holds also after smearing,
\begin{equation}
{\mbox tr}{\cal L}^L_\sigma
=
\sum_p L_p \sum_{r=1}^{\infty} w_p^r \frac{\delta_\sigma(L-rL_p)}
{\Abs{det(1-M_p^r)}}=
\frac{1}{2\pi i}
\int_{a-i\infty}^{a+i\infty} e^{sL}\frac{Z_w'(s)}{Z_w(s)}e^{(is\sigma)^2/2}ds
\label{eqn:trsmearid}
\end{equation}
where $\delta_{\sigma}(.)$ are gaussians of standard deviation $\sigma$.
We will be interested in very gross features of
periodic sums. 
Such
information is encoded in the behavior of the zeta function for small
$s$ and can be estimated
by the Baladi-Eckmann-Ruelle (BER) approximation\cite{BER}.
The probabilistic approach that underlies this
approximation has a long history, see eg \cite{prob}.
It was put into the context of Ruelle resonances 
by Baladi,Eckmann and Ruelle in \cite{BER}.
In \cite{PDreson,PDsin,PDzak,PDlyap} the formalism was generalized to include
zeta functions with general
thermodynamic weights, and the evaluation of chaotic averages.

Intermittent systems fluctuate between chaotic and quasi regular behavior.
The basic idea is
to define a surface of section such that all trajectories from the 
section back to itself traverse a chaotic region at least once. The coordinate 
on this surface of section will be denoted $x_s$.

Let $\Delta_s(x_s)$ the length of a trajectory starting at $x_s$
and back to the surface of section.
Further, let $w(x_s)$ be the weight integrated along this segment.
We define the (weighted) distribution of recurrence time as
\begin{equation}
p_w(L)=\int w(x_s) \delta(L-\Delta_s(x_s))dx_s  \  \  ,
\end{equation}
where we have assumed the measure $dx_s$ to be normalized: $\int dx_s =1$.
The approximate zeta function
is then given in terms of the Laplace transform
of this function
\begin{equation}
Z(s)\approx \hat{Z}(s) =1-\int p_w(L) e^{-sL}dL  \  \  .
\label{eqn:ZBER}
\end{equation}

\subsection{Penumbra diffraction}

\label{s:penum}

\begin{figure}
\epsfxsize=10cm
\epsfbox{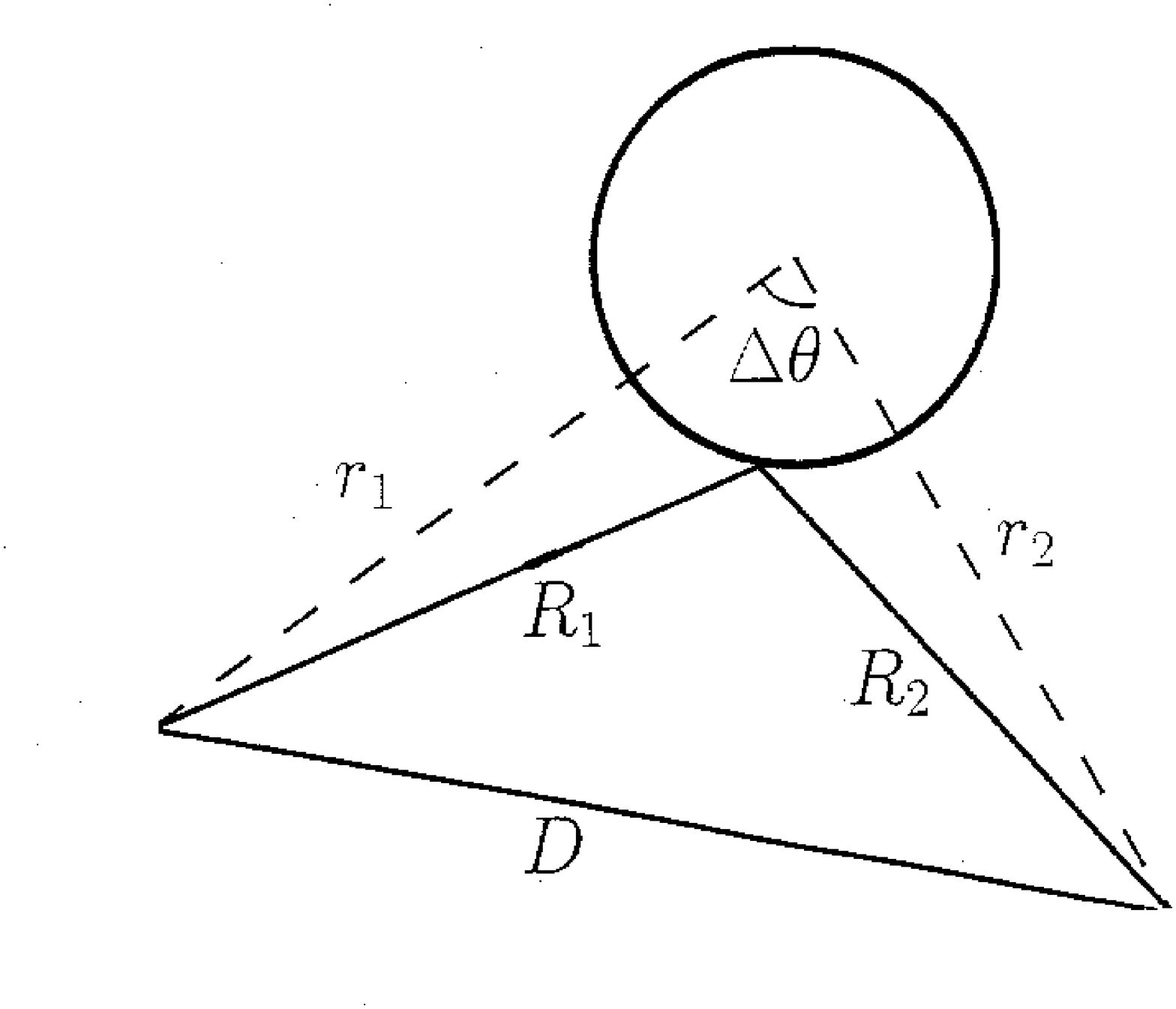}
\caption{Some notations used in the discussion of the one-disk Green
function.}
\label{f:naka3}
\end{figure}

A convenient starting point for deriving the semiclassical 
trace formula for billiards,
and to study its limitations, is the Boundary Integral Method \cite{BIM,Prim95,Boas}. 
The eigenvalues of the problem
are those for which the following integral equation has a solution
\begin{equation}
u({\bf r}(s))=2\int_S \frac{\partial G}{\partial \hat{n}_s}
({\bf r}(s),{\bf r}(s'))u({\bf r}(s'))ds' \label{eqn:bim}  \ \ .
\end{equation}
The integral is performed along the boundary of the billiard.
The function $u({\bf r}(s))$ is related to the wave function 
according to the normal derivative
\begin{equation}
u({\bf r}(s))=\frac{\partial \Psi({\bf r}(s))}{\partial \hat{n}_s} \ \ .
\end{equation}
The boundary ${\bf r}(s)$ is parameterized by the
Birkhoff coordinate $s$.

We can write eq \EqRef{bim} symbolically as a matrix equation
\begin{equation}
({\bf I}-{\bf A}){\bf U}=0 \ \ ,
\end{equation}
having a solution  when $\det({\bf I}-{\bf A})=0$.
We can write this determinant as
\begin{equation}
 \det ({\bf I}-{\bf A})=e^{ \mbox{\footnotesize  tr}\; \log ({\bf I}-{\bf A}) }
=e^{-\sum_{n=1}^{\infty} \frac{1}{n}\mbox{\footnotesize tr} ({\bf A}^n)} 
\  \  ,
\end{equation}
where
\begin{equation}
\begin{array}{l}
\mbox{tr} ({\bf A}^n)=\\
\; \; 2^n \int ds_1 \ldots ds_n 
\frac{\partial G}{\partial \hat{n}_{s_1}}({\bf r}(s_1),{\bf r}(s_2))
\ldots
\frac{\partial G}{\partial \hat{n}_{s_{n-1}}}({\bf r}(s_{n-1}),{\bf r}(s_n))
\frac{\partial G}{\partial \hat{n}_{s_n}}({\bf r}(s_n),{\bf r}(s_1)) 
\ \ . \end{array}
\label{eqn:traceAn}
\end{equation}

There is considerable liberty of choosing the Green function
$G({\bf r},{\bf r}')$.
In order to study the problem of Penumbra diffraction
Primack et.al \cite{Prim95}
suggested
to use the one  disk Green function (see below)
in \EqRef{bim}. The integral in \EqRef{bim} need then be performed only along the
square boundary.

The one disk Green function reads \cite{onedisk}
\begin{equation}
G(r_1,r_2,\Delta\theta)=
\frac{i}{8}\sum_{\ell =-\infty}^{\infty}
\left(H_\ell^{(2)}(kr_1)+ S_\ell(kR) H_\ell^{(1)}(kr_1)\right)
H_\ell^{(1)}(kr_2) e^{i\ell(\Delta\theta)} \ \ ,
\end{equation}
where $H_\ell^{(1)}(z)$ and $H_\ell^{(2)}(z)$ are Hankel functions
and $r_1$, $r_2$ and $\Delta \theta$ are explained in fig.\ \ref{f:naka3}. 
The phase shift function $S_\ell(kR)$ is defined by
\begin{equation}
S_\ell(kR) = -\frac{H_\ell^{(2)}(kR)}{H_\ell^{(1)}(kR)} \ \ .
\end{equation}

Using Poisson resummation we get
\begin{equation}
G(r_1,r_2,\Delta\theta)=\sum_{m=-\infty}^{\infty}G^{(m)}(r_1,r_2,\Delta\theta) \ \ ,
\end{equation}
where
\begin{equation}
G^{(m)}(r_1,r_2,\Delta\theta)=\frac{i}{8}\int_{-\infty}^{\infty}
\left(H_\ell^{(2)}(kr_1)+ S_\ell(kR) H_\ell^{(1)}(kr_1)\right)
H_\ell^{(1)}(kr_2) e^{i\ell(\Delta\theta+2\pi m)}d\ell  \label{eqn:Gmint} \ \ .
\end{equation}
The standard semiclassical result is obtained if we 
\begin{enumerate}
\item Retain only the integral approximation $G^{(m=0)}$ in \EqRef{Gmint} for the Green function,
semiclassically this means that classically forbidden orbits such as
creeping orbits are neglected.
\item Use the Debye approximations for the Hankel functions.
\item Compute the integrals by stationary phase. The term $\mbox{tr} ({\bf A}^n)$
in \EqRef{traceAn} will now contain contributions from all periodic orbits
with $n$ bounces on the square boundary. 
\end{enumerate}

\begin{figure}
\epsfxsize=11cm
\epsfbox{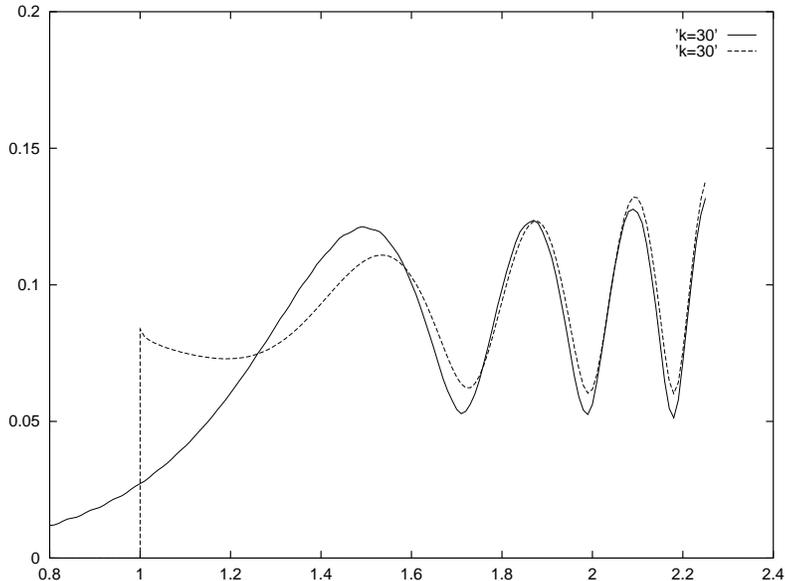}
\caption{The circle Green function versus impact parameter $d$
for fixed values of $k=30$, $R=1$ and $r_1=r_2=3$.
The full line represents the exact result and the dashed line 
the semiclassical approximation. }
\label{f:circle}
\end{figure}

In fig.\ \ref{f:circle}
we plot the circle Green function
together with its semiclassical limit.
Semiclassically there is a discontinuity at $d=R$ (where $d$ is the classical
impact parameter) 
marking the transition between the
lit region and the shadow.
In the exact Green function there is of course a smooth transition.
The interesting thing is that the exact result is suppressed as compared to
the semiclassical within a distance $d_{crit}$,
a destructive interference 
between the rays
starts already in the lit region and continues into the classical shadow.

This twilight zone was called Penumbra in \cite{Prim95}.
In appendix A we show that
\begin{equation}
d_{crit}= (1+\epsilon_{max}(kR) )R  \label{eqn:dcrit}  \  \  ,
\end{equation}
with
\begin{equation}
\epsilon_{max}(kR)=c  (kR)^{-2/3} \  \  ,   \label{eqn:epsdef}
\end{equation}
where $c$ depend only weakly (logarithmically) on $kR$, $r_1$, $r_2$ and $R$.
The subscripts $max$ was used in the appendix but will be dropped from now on.

In the penumbra the usually semiclassical look of the Green function is lost.
It is hard to foresee how what happens when this object is convoluted
with itself a la eq. \EqRef{traceAn}, in order to get contributions from
periodic orbits doing several passages 
of the penumbra.
To use the
periodic orbit apparatus described in sec \ref {s:traces} we need a
weight that is multiplicative along the flow.
However, we will not use the cycles to actually compute the spectrum,
we are only interested in estimating the error 
induced by penumbra diffraction, so we can
suffice with a rather crude weight.
Before we make up our mind how this should be achieved, let us discuss the
periodic orbits in the Sinai billiard.

The cycles can be coded by associating a (coprime)  lattice vector {\bf q}
with each disk-to-disk segment, see sec. \ref{s:implBER}.
In the limit of small $R$ any such periodic sequence 
$p_t=\overline{{\bf q}_1{\bf q}_2\ldots {\bf q}_n}$ can be realized in the system,
except for the rule that two consecutive lattice vector may not
be identical\cite{PDsin}.
Suppose now that we increase the size $R$ of the disk. Some segment of the 
periodic orbit, say ${\bf q}_1$, would then eventually need to go
through the disk which of course is prohibited, the cycle is then 
said to be pruned. Let's say that this happen when $R=R_{bif}$.
For $R<R_{bif}$,
there is another  cycle
$p_r=\overline{{\bf q}_{1a}{\bf q}_{1b}{\bf q}_2\ldots {\bf q}_n}$ that
actually do scatter at the disk when the companion $p_t$ just pass by.
When $R=R_{bif}$,  $p_t$ and $p_r$ overlap exactly and when $R>R_{bif}$ they are both
pruned.
This is an analogue of the saddle-node bifurcation in
smooth potential, and the only source or pruning in the Sinai billiard.
For a lucid discussion of pruning in billiards, see \cite{Kai}.

Obviously there is a close connection between pruning and penumbra diffraction.
The semiclassical weights drops suddenly to zero when the pair is pruned.
We have learned from our studies of the circle Green function that the
quantum pruning is more gradual.


To estimate the error we will say the pair effectively annihilate each other if
they are within the transition region discussed above.
Not that since $|\Lambda_{p_r}| \gg |\Lambda_{p_t}|$, we need in practice
only consider the removal of $p_t$.

So we define an cycle to be diffractive
if it passes the disk within the distance $\epsilon R$ 
(as given by \EqRef{epsdef}).
The error of the semiclassical weight $C_p$ 
for a diffractive cycle is thus defined as
$\delta C_p =-C_p$. A pseudo orbit is diffractive if
at least on of the participating prime cycles is diffractive. The error of the
diffractive pseudo orbit's 
amplitude is thus $\delta C_\alpha =-C_\alpha$

To exclude diffractive cycles from cycle sums we
introduce a multiplicative cycle weight $w_p$ 
such that $w_p=0$ if the orbit is affected by diffraction and
$w_p=1$ if not. Exactly,
how this is done is discussed in section \ref{s:prep}.
The associated pseudo cycle weight is
\begin{equation}
w_\alpha=\prod_p w_p^{n_p} \  \  .
\end{equation}
(with the convention that $0^0=1$)
where $n_p$ is restricted to $n_p \leq 1$.

One can also consider families of cycles with $n$ passages of the penumbra.
For each passage the trajectory can either choose to bounce off the disk or not.
So such a family thus consists of $2^n$ members and only one of them is unremoved 
from the calculation according to our
rule above. This is the one bouncing at every passage of the penumbra, the semiclassical
weight of this one is of course very much suppressed and the neglect of this orbit
is completely neglible.

The weight now depends on the parameter $\epsilon$.
We will skip the index $w$ on the classical zeta function
\EqRef{Zw} and instead denote it $Z(s;\epsilon)$.
Traces will be denoted
$\mbox{tr} {\cal L}_\epsilon ^L$ etc.

\section{Preparation}
\label{s:prep}

\subsection{Perturbation of the Berry-Keating zeros}

Let $k_0$ be a zero of $D(k)$ as given by \EqRef{BK} (with superscript omitted):
\begin{equation}
D(k_0)=2\sum_{\alpha} C_\alpha \cos (L_\alpha k-i\pi \bar{N}(k) ) =0  
\  \  .
\end{equation}
All pseudo orbit sums are hence forth subject to the
cutoff $L_\alpha<L_{BK}$, 
which will not be explicitly  written out in the sums.

We are interested in how  small errors in  the amplitudes $C_\alpha$ will
effect the location of this zero.
We thus add a small perturbation
\begin{equation}
\delta D(k) = 
2\sum_{\alpha} \delta C_\alpha \cos (L_\alpha k-i\pi \bar{N}(k) )
\  \  ,
\end{equation}
where $\delta C_\alpha$is the error of $C_\alpha$, 
and try to solve
\begin{equation}
D(k)+\delta D (k)=0  \  \  .
\end{equation}
We then expand
$k=k_0 +\delta k$ and consider the solution to
\begin{equation}
D'(k_0) \delta k+ \delta D(k_0)+\delta D'(k_0) 
\delta k=0  \  \  .
\end{equation}
We neglect the last term, it provide higher order corrections and get
\begin{equation}
\delta k=-\frac{\delta D(k_0)}{D'(k_0)}  \  \  .
\end{equation}
We will consider the perturbation of a typical zero, sitting at a distance
$\sim \bar{d}^{-1}$ (where $\bar{d}=\frac{d\bar{N}}{dk}$) 
from the neighboring zeros. If we assume that the 
oscillations of $D(k)$
are sine-like we can
relate the derivative at a typical zero $D'(k_0)$ to the mean square of the determinant
\begin{equation}
D'(k_0) =  \bar{d} \sqrt{2\pi^2
<D(k_0)^2>}  \  \   .
\end{equation}
We then get for the mean square of the shift
\begin{equation}
\expct{
\delta k^2}^{1/2}= \bar{d}^{-1}\left( 
\frac{\expct{\delta D(k_0)^2}}{2\pi^2<D(k_0)^2>}
\right)^{1/2}  \label{eqn:dk2}  \   \   .
\end{equation}
First we focus on the denominator of eq. \EqRef{dk2}.
We obtain 
\begin{equation}
<D(k_0)^2>=\frac{1}{2} 2^2 \sum_\alpha C_\alpha^2  \label{eqn:D2}   \  \  ,
\end{equation}
assuming that cross terms cancel out, cf section \ref{assum:diag}.

The average perturbation is, by the same arguments, given by
\begin{equation}
<\delta D^2 (k_0) >=2 \sum_\alpha \delta C_\alpha^2  \label{eqn:dD2}  \  \  ,
\end{equation}
and
\begin{equation}
 <\delta k^2>^{1/2}= \left(
\frac{\sum_\alpha \delta C_\alpha^2}{\sum_\alpha  C_\alpha^2}
\right)^{1/2}
\bar{d}^{-1}/\sqrt{2\pi^2} \equiv F(k_0 ) \bar{d}^{-1}/\sqrt{2\pi^2}
\ \   .
\end{equation}

Following our reasoning in sec \ref{s:penum} we put 
\begin{equation}
\delta C_\alpha^2= (1-w_\alpha) C_\alpha^2 \  \   .
\end{equation}
Recall that $w_\alpha$ is either $0$ or $1$.
We can rewrite the function $F$ as
\begin{equation}
F(k_0 ) = 
 \left(1-\frac{\sum_{\alpha} w_\alpha  C_\alpha^2}
{\sum_{\alpha }  C_\alpha^2}\right)^{1/2}   \ \   .\label{eqn:Fk0}
\end{equation}
Recall that we are only considering pseudo orbits such that $n_p \leq 1$ so
\begin{equation}
|C_{\alpha=p_1 p_2 \ldots p_k}|^2= |\Lambda_\alpha|^{-1} \equiv
|\Lambda_{p_1}\cdot \Lambda_{p_1} \ldots\Lambda_{p_k}|^{-1}
\  \  .
\end{equation}

We can now rewrite the numerator and denominator of \EqRef{Fk0}
in terms of the classical
weights $a(\epsilon)$ introduced in section \ref{s:traces}
\begin{eqnarray}
\sum_{\alpha} w_\alpha(\epsilon)  C_\alpha^2 = 
\sum_\alpha |a_\alpha(\epsilon)|    \label{eqn:ps1} \\
\sum_{\alpha}   C_\alpha^2  = \sum_\alpha |a_\alpha(0)|  \  \   .    
\label{eqn:ps2}
\end{eqnarray}

The perturbative approach taken in this section is only valid if
the predicted value of $F$ is small.
If $F$ is approaches unity there is no other interpretation than a failure
of the Berry-Keating formula to resolve individual states.

\subsection{Treating the pseudo orbit sums}

\label{s:treat}

The goal of this section is to relate the pseudo orbit sums
\EqRef{ps1} and \EqRef{ps2} to various zeta functions. 
There is an important distinction between these pseudo orbits sums and the
cycle expansion
\EqRef{clcycleexp} - the occurrence of the absolute values in 
eqs. \EqRef{ps1} and \EqRef{ps2}. To deal with this we will borrow a trick
from ref.\ \cite{AAC90}.

Consider the classical
zeta function (that is eq. \EqRef{Zw} with higher $m$ factors omitted)
\begin{equation}
Z(s;\epsilon)=\prod_p (1-w_p(\epsilon) \frac{e^{-sL_p}}{|\Lambda_p|}) \ \  .
\end{equation}

Consider the cycle expansion of a derived zeta function
(note the plus-sign!)
\begin{equation}
Z^+(s;\epsilon)=\prod_p (1+w_p(\epsilon) \frac{e^{-sL_p}}{|\Lambda_p|})=
\sum_\alpha b_\alpha(\epsilon) e^{-s L_\alpha}
\end{equation}
All $b_\alpha$ are positive, in fact, 
they are the summands
of eqs \EqRef{ps1} and \EqRef{ps1}
\begin{equation}
b_\alpha(\epsilon)= |a_\alpha(\epsilon)|  \  \    ,
\end{equation}

We rewrite this as 
\begin{equation}
Z^+(s;\epsilon)=
\int_{0}^{\infty} b(L;\epsilon) e^{-sL}\; dL
\end{equation}
where
\begin{equation}
b(L;\epsilon)=\sum_\alpha b_\alpha \; \delta(L-L_\alpha)  \  \  .
\end{equation}
Conversely, the function $b(L;\epsilon)$ is related to the zeta function
by means of an inverse Laplace transform
\begin{equation}
b(L;\epsilon)= \frac{1}{2\pi i} \int_{\sigma-i\infty}^{\sigma+i\infty}Z^+(s;\epsilon)
e^{sL}ds \ \ . \label{eqn:b}
\end{equation}

The semiclassical error can now be written in terms of the function
$b(L;\epsilon)$
\begin{equation}
F=\sqrt{1-\frac{\int_0^{L_{BK}} b(L;\epsilon) dL }
{\int_0^{L_{BK}} b(L;0) dL}}  \label{eqn:Fb}
\end{equation}
where $L_{BK}$ is given by
eq. \EqRef{LBK} and $\epsilon$ is given eq. \EqRef{epsdef}.

If we now define yet another zeta function
\begin{equation}
Z_2(s;\epsilon) =\prod_p (1-w_p^2 \frac{e^{-2sL_p}}{|\Lambda_p|^2})
\label{eqn:Z2def}
\end{equation}
we can rewrite 
$Z^+(s;\epsilon)$ as
\begin{equation}
Z^+(s;\epsilon)= \frac{Z_2(s;\epsilon)}{Z(s;\epsilon)}  \  \   .
\end{equation}
If the involved zeta functions are entire we  can write
\begin{equation}
b(L;\epsilon) = \sum_\gamma \mbox{res} \frac{Z_2(s;\epsilon )}{Z(s;\epsilon )}
\mid_{s=s_\gamma}
e^{s_\gamma L}
\end{equation}
which should be compared with \EqRef{trsgamma}.
The asymptotic behavior of $b(L;\epsilon)$ 
is (under much milder assumptions on the analytic properties) related
to the leading zero
$s_0(\epsilon)$ 
\begin{equation}
b(L;\epsilon)
\sim \mbox{res}\frac{Z_2(s;\epsilon)}
{Z(s;\epsilon)}\mid_{s=s_0} e^{s_0 L}= 
\frac{Z_2(s_0(\epsilon);\epsilon)}
{Z'(s_0(\epsilon);\epsilon)} e^{s_0 L}  \label{eqn:bas}
\end{equation}
This leading asymptotic behavior is all we need to evaluate the integrals
in eq.\ \EqRef{Fb} since we are interested in the asymptotics of
the function $F$. Moreover, as discussed in section \ref{s:BERtheory},
the behavior of zeta functions close to the origin is insensitive to fine details
in the spectrum of periodic orbits, a theory for the large scale structure
of periodic orbits exist and will be worked out in detail in the next section.

\subsection{Implementing the BER approximation
for the Sinai billiard}

\label{s:implBER}

The BER approximation is very well suited for the Sinai billiard, 
in particular if the scatterer
is small.
The obvious choice of the surface of section 
is provided by the disk itself \cite{PDsin}.

Consider now the unfolded 
representation of the Sinai billiard. A trajectory segment from the disk
to itself can thus be considered as going from one disk, associated
with lattice vector $(0,0)$  to some other disk represented by
lattice vector {\bf q}.
All segments going to {\bf q} have essentially the same length
$q=|{\bf q}|$ and the following simple expression
for the distribution of recurrence lengths
will suffice for our purposes.
\begin{equation}
p(L;\epsilon)=\sum_{\bf q} \hat{a}_{\bf q}(\epsilon)
 \delta(L-q)   \label{eqn:p_level1}
\  \  ,
\end{equation}
where
\begin{equation}
\hat{a}_{\bf q}(\epsilon)=\int_{\Omega_q} w(x_s)dx_s \label{eqn:a_level1} 
\  \  .
\end{equation}
$\Omega_q$ is the set of initial points $x_s$ whose target is
{\bf q}.
The approximate zeta function is then
\begin{equation}
\hat{Z}(s;\epsilon)=1-\int_0^\infty p(L;\epsilon) e^{-sL}ds=
1-\sum_{\bf q} \hat{a}_{\bf q}(\epsilon) e^{-sq} 
\  \  .  \label{eqn:Z_level1}
\end{equation}

The weight, as defined in sec. \ref{s:penum} will be zero, 
$w(x_s)=0$, if the trajectory 
starting at $x_s$ and heading for disk {\bf q} pass some other disk within
a distance $\epsilon R$ before actually hitting {\bf q},
otherwise it is equal one, $w(x_s)=1$.

If $\epsilon=0$, then $\hat{a}_{\bf q}$ is the phase space area corresponding to
disk {\bf q}.
  
The computation of
$p(L;\epsilon)$
will unfortunately be long and boring. It will  run in parallel
with ref \cite{PDsmall} in large parts and we will frequently refer to that paper.
All error estimates in \cite{PDsmall} carry over directly,
so we will simply omit them below to make things a little more transparent.
Nevertheless, we will without hesitation 
display equations as 
strict equalities,  but the reader
should bear in mind that all expressions are valid in the small $R$ limit.

\begin{figure}
\epsfxsize=10cm
\epsfbox{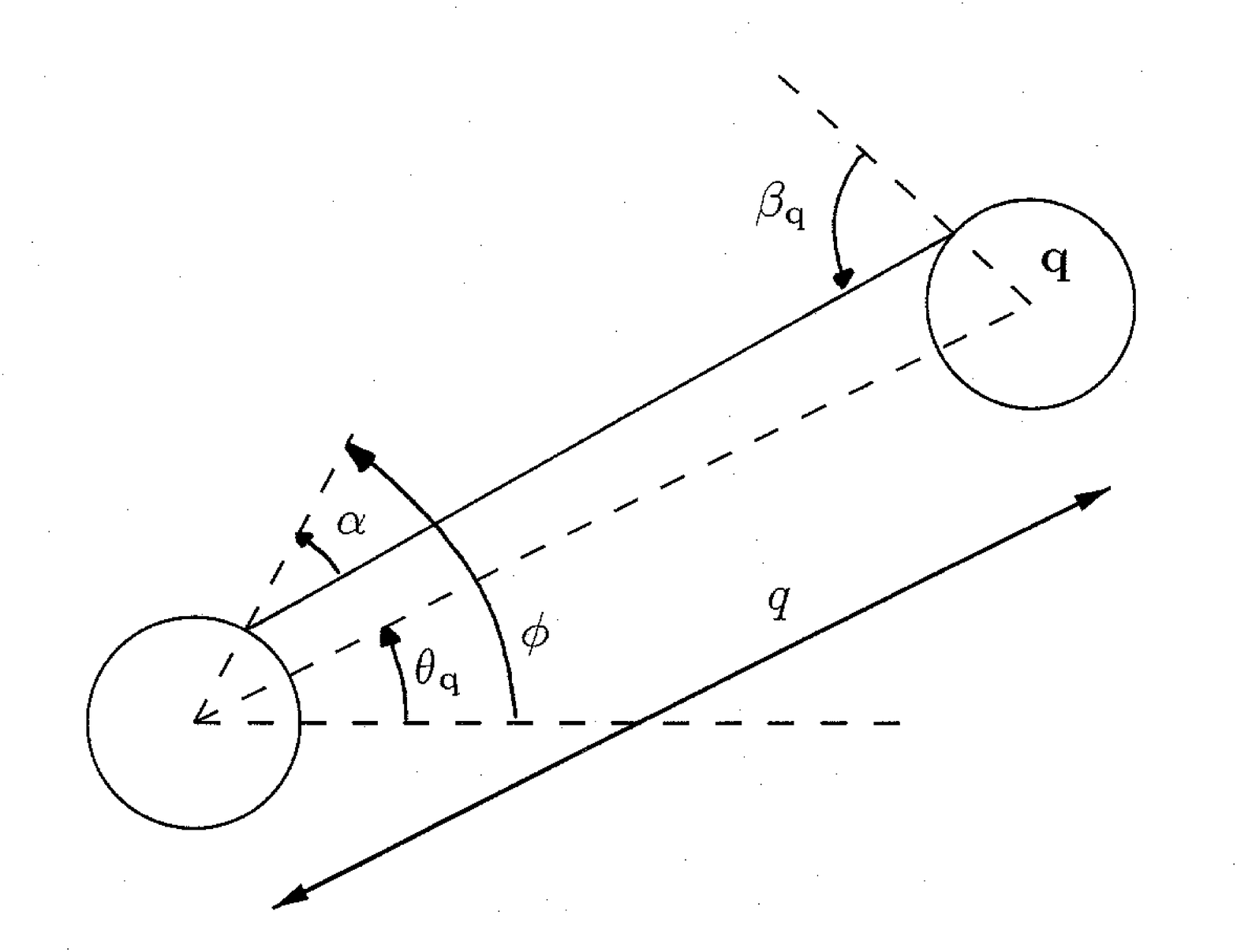}
\caption{Definition of various angles associated with scattering to disk {\bf q}.}
\label{f:small2}
\end{figure}

Consider now a trajectory hitting disk ${\bf q}$.
The relation between the phase space variables $\phi$ and $\alpha$,
see fig \ref{f:small2},
and the scattering angle $\beta_{\bf q}$ on disk ${\bf q}$ is given by 
\begin{equation}
R_{\bf q}\sin \beta_{\bf q}-R\sin\alpha =q\sin (\phi -\theta_{\bf q}-\alpha)
\  \  ,
\end{equation}
where $\theta_{\bf q}$ is the polar angle of the lattice vector ${\bf q}$,
and $R_{\bf q}$ is the radius of the target disk.
The argument of
$\sin (\phi - \theta_{\bf q}-\alpha)$ 
is small when ${\bf q}$ is large so we can expand the sine
and get 
\begin{equation}
R_{\bf q} \sin \beta_{\bf q}-R\sin\alpha =q (\phi -\theta_{\bf q}-\alpha)
\label{eqn:RRR}  \  \  .
\end{equation}

\begin{figure}
\epsfxsize=10cm
\epsfbox{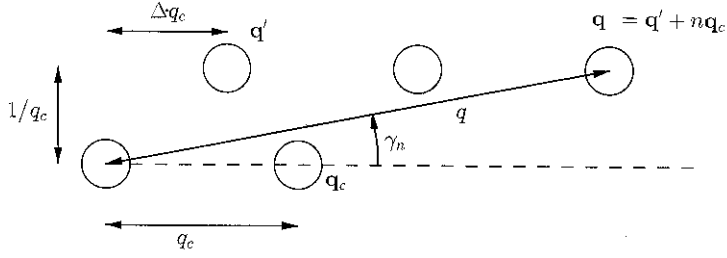}
\caption{Scattering to disk ${\bf q}={\bf q}'+n{\bf q}_c$ with $n=2$ in
the corridor ${\bf q}_c$.}
\label{f:small3}
\end{figure}

It is easy to see that only disks represented by coprime lattice vectors
are accessible.
In ref. \cite{PDsmall} we showed that any coprime lattice vector {\bf q} can uniquely
be written on the form
${\bf q}={\bf q}'+n{\bf q}_c$, where $n\geq 2$ and
${\bf q}'$ represents the lattice points closest to
the line from $(0,0)$ to ${\bf q}_c$.
We say that disk ${\bf q}$ lies in the ${\bf q}_c$ {\em corridor}, see 
fig \ref{f:small3}.

Actually there are two such neighboring lattice points
for each corridor vector ${\bf q}_c$,
one with smaller and one with larger polar angle.
Below we assume that ${\bf q}'$ is the one with the larger polar angle, the
other case is completely analogous,
and is accounted for by multiplying 
by a factor of two on some strategic occasions,
see below.

The disk under observation, ${\bf q}$, is potentially
shadowed only by two disks, namely
 ${\bf q}-{\bf q}_c$
and ${\bf q}_c$, see fig. \ref{f:small3}.
We implement the weight $w(x_s)$ by 
simply inflating
these two disks  from radius $R$ to $(1+\epsilon)R$.

To see how $\Omega_{{\bf q}_c}$ shadows $\Omega_{\bf q}$
we replace {\bf q} in \EqRef{RRR} by ${\bf q}_c$ and put
\begin{equation}
\begin{array}{l}
\beta_{{\bf q}_c}=\pi/2\\
R_{{\bf q}_c}=(1+\epsilon)R\\
\theta_{{\bf q}_c}=\theta_{\bf q}-1/(q\;q_c)
\end{array}  \  \  .
\end{equation}
This gives an equation (in terms of  phase space variables 
$\alpha$ and $\phi$) for the relevant boundary of $\Omega_{{\bf q}_c}$
\begin{equation}
(1+\epsilon)R-R\sin \alpha =q_c ( (\phi - \theta_{\bf q}+
\frac{1}{q\;q_c}-\alpha))  \label{eqn:ett}  \  \  .
\end{equation}

We can now treat $\Omega_{{\bf q}-{\bf q}_c}$ in the same way.
We replace {\bf q} in \EqRef{RRR} by ${\bf q}-{\bf q}_c$ and put
\begin{equation}
\begin{array}{l}
\beta_{{\bf q}-{\bf q}_c}=-\pi/2\\
R_{{\bf q}-{\bf q}_c}=(1+\epsilon)R\\
\theta_{{\bf q}-{\bf q}_c}=\theta_{\bf q}+1/(q\;q_c)
\end{array}  \  \  ,
\end{equation}
and get the equation for the borderline of $\Omega_{{\bf q}-{\bf q}_c}$
\begin{equation}
-(1+\epsilon)R-R\sin \alpha =(q -q_c) ( (\phi - \theta_{\bf q}-
\frac{1}{q\;q_c}-\alpha))  \  \  .  \label{eqn:tvaa}
\end{equation}

Next we want to change variables from ($ \alpha, \phi)$ to
($\sin\alpha,\sin\beta_{\bf q})$),the relation is given by eq \EqRef{RRR}.

By combining equation \EqRef{RRR} and \EqRef{ett}
we see that
the shadowing of $\Omega_{{\bf q}_c}$ corresponds to the straight 
line
\begin{equation}
q_c\sin\beta_{\bf q} +(q_n-q_c)\sin \alpha =q(1+\epsilon)-\frac{1}{R}
  \ \ .  \label{eqn:two}
\end{equation}
Similarly,
by combining equation \EqRef{RRR} and \EqRef{tvaa}
we see that
the shadowing of $\Omega_{{\bf q}-{\bf q}_c}$
is given by
\begin{equation} 
(q-q_c)\sin \beta_{\bf q} +q_c \sin \alpha=q(1+\epsilon)-\frac{1}{R} \ \ .
\label{eqn:three}
\end{equation}
If {\bf q} were not shadowed, $\Omega_{\bf q}$ would be given by
$-1<\sin \beta_{\bf q} <1$ and $-1<\sin \alpha <1$.
So the integral \EqRef{a_level1} is simply  the area 
of the remainder of this square,
lying inside the lines given by eqs. \EqRef{two} and \EqRef{three}.

The integration element $dx_s$ in \EqRef{a_level1}
over $\Omega_{\bf q}$
is (to leading order) 
\begin{equation}
dx_s=\frac{ R}{4\pi q}d( \sin \alpha)d( \sin  
\beta_{\bf q})  \ \ . 
\end{equation}
It is normalized in such a way that the integral over one octant of the
plane is unity.

So we arrive at the following results 
\begin{equation}
\hat{a}_{\bf q}(\epsilon)=\frac{R}{\pi q}
\cdot \left\{ \begin{array}{lrll}
1 & 0&<q<&\frac{1}{2R}\frac{1}{1+\epsilon/2} \\
(1-\frac{(1/2R-q-\epsilon q/2)^2}{q_c(q-q_c)}) & 
\frac{1}{2R}\frac{1}{1+\epsilon/}&<q<&(\frac{1}{2R}+q_c)\frac{1}{1+\epsilon/2}\\
\frac{(1/2R-q_c-\epsilon q/2)^2}{(q-q_c)(q-2q_c)} & (\frac{1}{2R}+q_c)\frac{1}{1+\epsilon/2}
 &<q<&\frac{2}{\epsilon} (\frac{1}{2R}-q_c)\\
0 & \frac{2}{\epsilon} (\frac{1}{2R}-q_c) &<q&
\end{array} \right. \label{eqn:a_level2}
\end{equation}
The result can be described in the following way.
All disks inside a radius $q<\frac{1}{2R}\frac{1}{1+\epsilon/2}$
are unshadowed and the corresponding trajectory segments not diffractive.
Outside this horizon the accessible disks are aligned along corridors,
each corridor characterized by the
vector ${\bf q}_c$, subject to the condition
$q_c<\frac{1}{2R}\frac{1}{1+\epsilon/2}$.
Segments longer than $\frac{2}{\epsilon} (\frac{1}{2R}-q_c)$ in a particular
corridor ${\bf q}_c$ are always affected by diffraction. 
Segments longer than $\frac{2}{\epsilon} \frac{1}{2R}$ are always
diffractive.

This means that there is a one to one correspondence between corridors 
(beyond the horizon)
and accessible
disks inside the horizon, so in order to
perform the sum \EqRef{Z_level1} we just need to know the finite number of 
coprime lattice points
inside the horizon,
exactly how the sum should administrated we be clear below.

\begin{figure}
\epsfxsize=11cm
\epsfbox{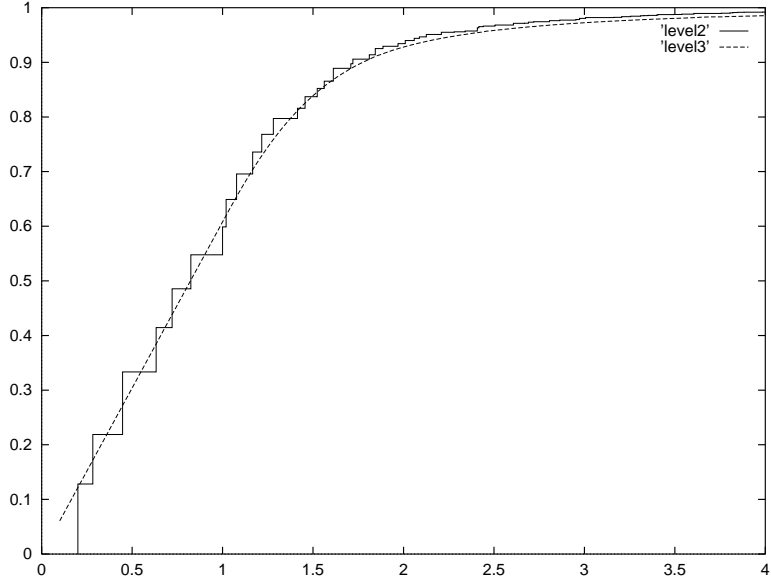}
\caption{The integral of $p(L(\xi))$ versus $\xi$. The radius is $R=0.1$ and 
$\epsilon=0$. The (full) staircase curve is obtained from eq.\ \EqRef{p_level1}
with amplitudes $\hat{a}_{\bf q}(\epsilon)$ given by \EqRef{a_level2}.
The (dashed) smooth curve is obtained from eq.\ 
(\ref{eqn:f1_level3},\ref{eqn:f2_level3},\ref{eqn:f3_level3}).}
\label{f:level23}
\end{figure}

\begin{figure}
\epsfxsize=11cm
\epsfbox{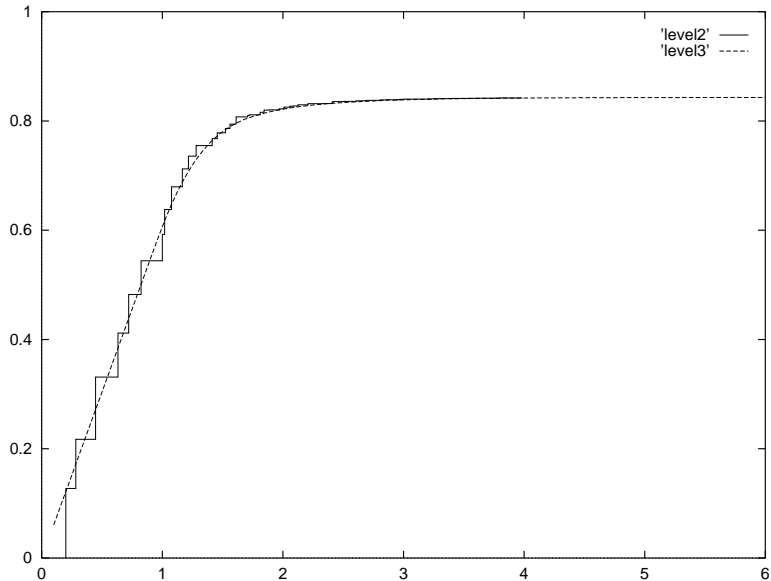}
\caption{Same as fig.\ \ref{f:level23} but with 
$\epsilon=0.2$.}
\label{f:level23eps}
\end{figure}


We can now obtain an approximate zeta function by plugging
these $\hat{a}_q$'s into \EqRef{Z_level1}.
The integrated recurrence time distribution 
$\int_0^L p(L')dL'=\sum_{{\bf q}:|{\bf q}|<L}\hat{a}_{\bf q}$ 
is plotted in figs.\ \ref{f:level23} and \ref{f:level23eps}.

However, if $R$ is small, 
there is a vast number of of coprime lattice points inside the horizon, and
according to number theory, they tend to be uniformly distributed over 
the plane.
As a matter of fact, one can use this uniformity to turn the sum into an
integral and write down explicit formula for the distribution of recurrence 
times \EqRef{p_level1}.

Asymptotically, there are $6L^2/\pi$ coprime lattice points
{\bf q} such that $|{\bf q}|<L$ in the first octant so that
the mean density of coprime lattice points is in an asymptotic sense
\begin{equation}
d_c(L)=\frac{12L}{\pi}  \  \  .
\label{eqn:dcL}
\end{equation}

\vspace{0.4cm}
\noindent
{\it (i)} First we consider the case $L<\frac{1}{2R}\frac{1}{1+\epsilon/2}$.\\
Then $\hat{a}_{\bf q}$ is a function of $q$ only, 
cf. eq. \EqRef{a_level2}.
The distribution function $p(L;\epsilon)$ becomes
\[
p(L;\epsilon)=\sum_{\bf q} \hat{a}_{\bf q}  
\delta_\sigma (L-q)  =
\sum_{\bf q} \frac{R}{\pi q} 
\delta_\sigma (L-q) 
\]
\begin{equation}
= d_c(L) \frac{R}{\pi L}=
\frac{12R}{\pi^2}  \ \ .
\end{equation}
We introduce the rescaled length $\xi =2R L$ and rewrite
\begin{equation}
p(L(\xi);\epsilon)=
\frac{12R}{\pi^2} \hspace{0.5cm} \xi < \frac{1}{1+\epsilon/2} \ \ .
\end{equation}

\vspace{0.4cm}
\noindent
{\it (ii)} Next we consider the {\em transition region} $1/2R<L<1/R$.\\
According to eq. \EqRef{a_level2}
the amplitudes $\hat{a}_{\bf q}$
depend on the  size of corridor $q_c=|{\bf q}_c|$ and the length of {\bf q}: 
$\hat{a}_{\bf q}=\hat{a}(q_c,q)$.
The length $q$ is approximately $q\approx q'+nq_c\equiv
\Delta q_c+nq_c$, where $\Delta$ is a number such that $0<\Delta<1$,
see fig \ref{f:small3}. 
We get
\begin{equation}
p(L;\epsilon)=2\sum_{{\bf q}_c}\sum_{n=2}^{\infty}\hat{a}(q_c,q)
\delta_\sigma(L-q)
\end{equation}
\[
=2\sum_{{\bf q}_c}\sum_{n=2}^{\infty}\hat{a}(q_c,q)\delta_\sigma(L-q_cn-\Delta q_c)
  \ \ ,  
\]
where we inserted the factor 2 to account for both neighbors
of ${\bf q}_c$.

\begin{figure}
\epsfxsize=11cm
\epsfbox{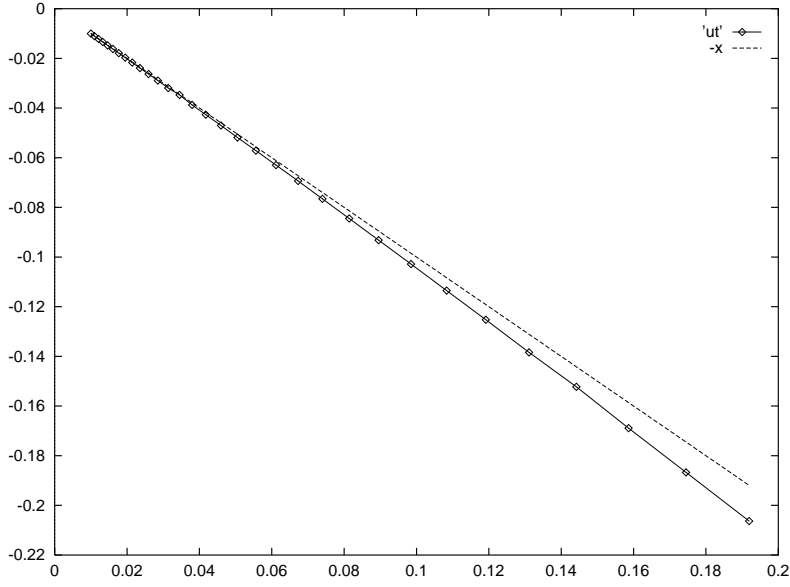}
\caption{The leading zero $z_0(\epsilon)$ versus $\epsilon$ compared
with the asymptotic formula \EqRef{z0_as}.}
\label{f:leading}
\end{figure}

We will now turn the sums over ${\bf q}_c$ and $n$ into an
integral over the density of coprime lattice vectors.
The parameter $\Delta$ is uniformly distributed in the interval
$0<\Delta<1$ \cite{PDsin}, this means that we can just integrate $n$
from $n=2$
to $\infty$.

\[
p(L;\epsilon)=2\int_0^{1/2R-\epsilon L/2} dq_c\; d_c(q_c) \; 
\hat{a}(q_c,L) \int_{n=2}^{\infty}
\delta_\sigma(L-q_cn) \ \ ,
\]
\begin{equation}
=2\int_0^{1/2R-\epsilon L/2} dq_c\; d_c(q_c) \; \hat{a}(q_c,L)\frac{1}{q_c} 
\theta_\sigma(L-2q_c)
\end{equation}
\[
=2\int_0^{\mbox{min}(1/2R-\epsilon L/2,L/2)} dq_c\; d_c(q_c) \; 
\hat{a}(q_c,L)\frac{1}{q_c}  \ \ ,
\]
where $\theta(x)$ is the unit step function. 

Since we are considering the region
$\frac{1}{2R}\frac{1}{1+\epsilon/2} < q < \frac{1}{R}\frac{1}{1+\epsilon}$, 
we have
$\mbox{min}(1/2R-\epsilon L/2,L/2)=L/2$.
Next we insert the expression for $\hat{a}(q_c,q)$ from eqs. \EqRef{a_level2} and
 $d_c(q_c)$ from eq. \EqRef{dcL}
\begin{equation}
p(L;\epsilon)=2\int_{l+\epsilon L/2-1/2R}^{l/2}dq_c \; \frac{12q_c}{\pi}\; \frac{1}{q_c}
\frac{R}{\pi L}(1-\frac{(1/2R-L-\epsilon L/2)^2}{q_c(L-q_c)})
\end{equation}
\[
+2\int_{0}^{L-\epsilon L/2-1/2R}dq_c\;\frac{12q_c}{\pi}\; \frac{1}{q_c}
\frac{R}{\pi L}\frac{(1/2R-q_c-\epsilon L/2)^2}{(L-q_c)(l-2q_c)}  \ \ .
\]
We change integration variable to $\eta =2Rq_c$ and use as before
$\xi =2R L$.
This leaves us the following integral to solve
\begin{equation}
p(L(\xi);\epsilon)=\frac{24R}{\pi^2} \left(
\int_{\xi+\epsilon\xi/2-1}^{\xi/2}
\frac{d\eta}{\xi}(1-\frac{(1-\xi-\epsilon\xi/2)^2}{\eta(\xi-\eta)})+
\int_0^{\xi+\epsilon\xi/2-1}
\frac{d\eta}{\xi}\frac{(1-\eta-\epsilon\xi/2)^2}{(\xi-\eta)(\xi-2\eta)}
\right)  \  \   .
\label{eqn:integral1}
\end{equation}
The result of this integral is displayed below \EqRef{f2_level3}.

\vspace{0.4cm}
\noindent
{\it (iii)}
Now remains only the case 
$\frac{1}{R}\frac{1}{1+\epsilon}<L<\frac{2}{\epsilon}$:\\
The calculation is completely analogous
to the previous case and we get
\[
p(L(\xi);\epsilon)=2\int_{0}^{1/2R-\epsilon L/2}dq_c \; 
d_c(q_c)\frac{1}{q_c}\;a(q_c,q=L)
\]
\begin{equation}
=2\int_{0}^{1/2R-\epsilon L/2}dq_c \; \frac{12q_c}{\pi}\; \frac{1}{q_c}
\frac{R}{\pi L}\frac{(1/2R-q_c-\epsilon L/2)^2}{(L-q_c)(L-2q_c)}
\label{eqn:integral2}
\end{equation}
\[
=\frac{24R}{\pi^2}
\int_0^{1-\epsilon\xi/2}\frac{d\eta}{\xi}
\frac{(1-\eta-\epsilon\xi/2)^2}{(\xi-\eta)(\xi-2\eta)}  \  \  .
\]

\begin{figure}
\epsfxsize=11cm
\epsfbox{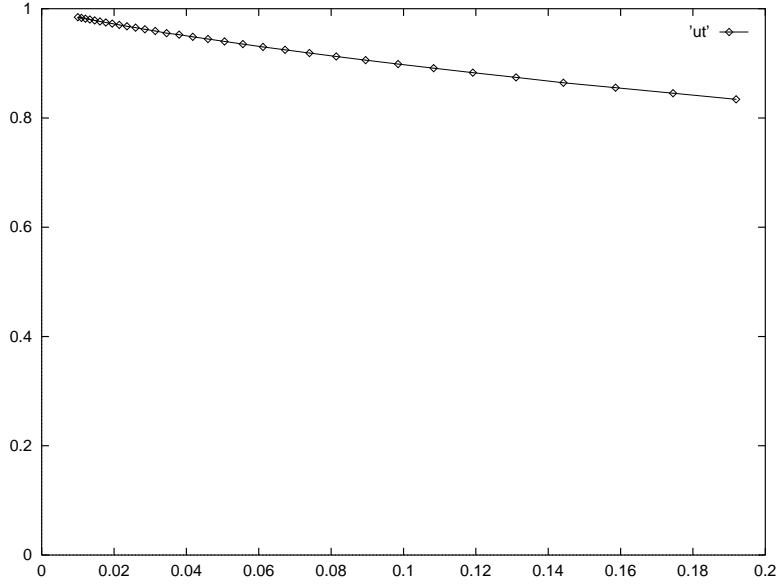}
\caption{The derivative $\hat{Z}'(z_0)$ 
evaluated at the leading zero versus $\epsilon$.
It approached unity as $\epsilon \rightarrow 0$ as predicted by eq.\ \EqRef{Zpz0}.}
\label{f:derivative}
\end{figure}

It is time to summarize the results.
It is natural to display the final results in terms of the distribution of the rescaled
recurrence lengths $\xi$ rather than $L$. We call the distribution $f(\xi;\epsilon)$ 
and it is 
trivially related
to $p(L;\epsilon)$ according to
\begin{equation}
p(L;\epsilon) dL = f(\xi;\epsilon) d\xi  \  \   .
\end{equation}
Inside the horizon we already have
\begin{equation}
\begin{array}{ll}
f(\xi;\epsilon)=\frac{6}{\pi^2} & \xi<\frac{1}{1+\epsilon/2} 
\end{array} \label{eqn:f1_level3}
\end{equation}
Beyond the horizon we get, after having performed the integrals
\EqRef{integral1} and \EqRef{integral2}
\footnote{The results of these two integrals
can be summarized in one formula, note the absolute value in one of the logarithms}
\begin{equation}
\begin{array}{lll}
f(\xi;\epsilon)=&
\frac{3}{\pi^2 \xi^2} \left( 2\xi-
\epsilon\xi^2 +[4\xi-3\xi^2-2\epsilon \xi^2] \log \xi + \right.        &  \\
 &  [4\xi^2-8\xi^2+4-4\epsilon\xi^2-\epsilon^2\xi^2-4\epsilon\xi] 
\log (\xi(1+\epsilon/2)-1)& \\ 
&+
\left.[4\xi-\xi^2-4-2\epsilon\xi^2-\epsilon^2\xi^2+4\epsilon\xi] 
\log |\xi(1+\epsilon)-2 |\right) & \frac{1}{1+\epsilon/2} < \xi <2/\epsilon
\end{array}  \label{eqn:f2_level3} \  \  ,
\end{equation}
This is valid up to the point $\xi=2/\epsilon$ where the function chokes.
After that we have
\begin{equation}
\begin{array}{ll}
f(\xi;\epsilon)= 0 & 2/\epsilon < \xi\end{array} \  \  .
\label{eqn:f3_level3}
\end{equation}
The integral  $\int_0^\xi f(\xi';\epsilon) d\xi'$ if these expressions is plotted in 
figs.\ \ref{f:level23} and \ref{f:level23eps}. We note that the
statistical treatment  of the lattice vectors
works surprisingly well, even for such a "large" radius as $R=0.1$.

It is also natural 
to let the zeta functions depend on a rescaled variable $z$ as defined  by
\begin{equation}
s=2Rz  \  \  .
\end{equation}

A power series expansion of the the zeta function is related to the moments of
the distribution $f(\xi;\epsilon)$
\begin{equation}
\hat{Z}(z;\epsilon)=1-\int e^{-z\xi}f(\xi;\epsilon)\; d\xi=
1-\sum_{m=0}^{\infty} \frac{(-z)^m}{m!} \int \xi^m f(\xi;\epsilon)\; d\xi
\  \  , \label{eqn:Zzpower}
\end{equation}
These moments can be computed from eqs.\ \EqRef{f1_level3},\EqRef{f2_level3} and 
\EqRef{f3_level3} and are found to be
\begin{equation}
\int \xi^m f(\xi;\epsilon)\; d\xi=\left\{
\begin{array}{ll}
1-\epsilon+O(\epsilon^2 \log \epsilon) & m=0\\
 (1+O(\epsilon \log \epsilon)) & m=1\\ 
 (O( \log \epsilon)) & m=2\\
 O(1/\epsilon^{m-2}) & m\geq 3 \end{array} \right. \label{eqn:moments}
\  \  .
\end{equation}

The leading zero
$z_0$ of the zeta function can be computed from eqs.\ \EqRef{Zzpower} and
\EqRef{moments},
and is found to be
\begin{equation}
z_0=-\epsilon+O(\epsilon^2 \log \epsilon)  \label{eqn:z0_as} \  \  .
\end{equation}

We are also interested in the derivative of the zeta function evaluated at at
this zero
\begin{equation}
\hat{Z}'(z_0)=1+O(\epsilon \log \epsilon )  \label{eqn:Zpz0}  \  \   .
\end{equation}
In figs.\ \ref{f:leading} and \ref{f:derivative} we compare these asymptotic
formulas with results from numerical computation of the BER zeta function as obtained
from eqs.\ \EqRef{f1_level3},\EqRef{f2_level3} and 
\EqRef{f3_level3}.

\begin{figure}
\epsfxsize=11cm
\epsfbox{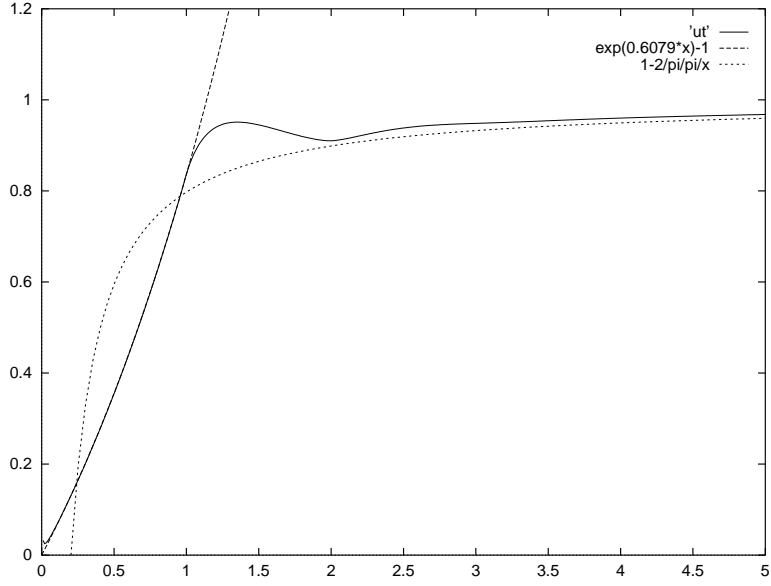}
\caption{The trace \EqRef{trace_level3} for $\epsilon=0$ (full line).
Comparison is made with eq.\ \EqRef{small_xi} for small values of $\xi$ (dashed line)
and eq.\ \EqRef{pow_corr} for large values of $\xi$ (dotted line).
}
\label{f:trace00}
\end{figure}

\begin{figure}
\epsfxsize=11cm
\epsfbox{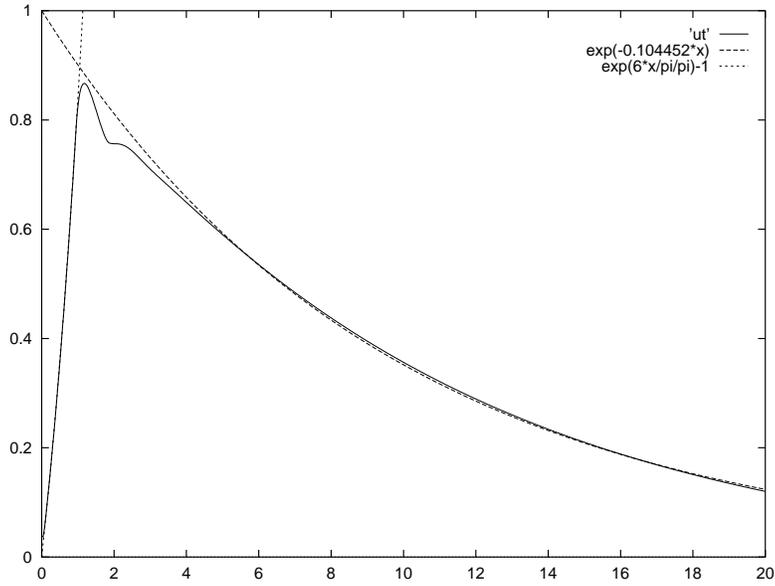}
\caption{The trace \EqRef{trace_level3} for $\epsilon=0.1$.
Comparison is made with eq.\ \EqRef{small_xi} for small values of $\xi$ (dotted line)
and eq.\ \EqRef{large_xi} for large values of $\xi$ (dashed line).}
\label{f:traceeps}
\end{figure}

Traces will not be a direct concern to us. But, by computing traces in the BER framework,
we can get an idea of the influence of non leading zeros by computing
the trace in the BER approximation. 
That information is relevant for the next section.
The computation is done numerically by FFT technique.

The trace formula is in our rescaled units
\begin{equation}
tr {\cal L}_\epsilon^\xi = \frac{1}{2\pi i}
\int_{a-i\infty}^{a+i\infty} e^{z\xi}\frac{d}{dz}\log \hat{Z}_w (z)dz   
\ \ .
\label{eqn:trace_level3}
\end{equation}
Traces for $\epsilon=0$ and $\epsilon=0.1$
are plotted in figs.\ \ref{f:trace00} and \ref{f:traceeps}.
For small $\xi$ the trace (in the BER approximation) is given by 
\begin{equation}
tr {\cal L}_\epsilon^\xi = \exp (6\xi/\pi^2) -1 \; \; \; 
\xi<\frac{1}{1+\epsilon/2}  \label{eqn:small_xi}  \  \  .
\end{equation}
If the reader want to verify this,
the following hint should be useful:
The trace $tr {\cal L}_\epsilon^\xi$ in the range $\xi < \Xi$ 
depends only on the behavior of the distribution
$f(\xi)$ in the same range  ($\xi < \Xi$) and $f$ is constant in this range. 

In the large $\xi$ limit, the asymptotic behavior is given by the leading
zero $z_0$, and thus
\begin{equation}
tr {\cal L}_\epsilon^\xi  \sim \exp (z_0 \xi)\sim  \exp (-\epsilon \xi)
\label{eqn:large_xi} \  \  ,
\end{equation}
see fig \ref{f:traceeps}.
We observe that this asymptotic result settles down very early, that is long
before the natural scale $2/\epsilon$, cf.\ eq. \EqRef{f3_level3}.

For any finite $\epsilon$ the zeta function is entire since $p(L)$ has compact
support. But when $\epsilon \rightarrow 0$ zeros will accumulate along
the negative real $z$- axis, building up a branch cut.
For the limiting case $\epsilon=0$ this will lead to a power law correction
\begin{equation}
tr {\cal L}^\xi  \sim1-\frac{2}{\pi^2\xi}
\label{eqn:pow_corr}  \  \  ,
\end{equation}
see fig \ref{f:trace00}. 
This power law will not be essential in the following.

\vspace{0.5cm}

We also need to know the value of $Z_2(z_0;\epsilon)$ as defined in 
\EqRef{Z2def}. 
This zeta function contain the square of $|\Lambda_p|$ in the denominator.
This means that at $z=z_0$ (which is close to the origin for small $\epsilon$)
the value of the zeta function
will be dominated by the shortest cycles, which, for small $\epsilon$, 
will be non diffractive.
This implies that  $Z_2(z_0;\epsilon)$
tends to a constant faster than $\hat{Z}'(z_0)$ do, as $\epsilon \rightarrow 0$.
A simple estimate, based on the methods in \cite{PDlyap} suggests that
\begin{equation}
Z_2(z_0,\epsilon)=1+O(R^2 \log R)+O(R^2)O(\epsilon)  \label{eqn:Z2_slow}
\end{equation}
which should be compared with eq.\ \EqRef{Zpz0}.

\section{Results}

\label{s:result}

We now possess all the tools we need to finally be able to 
compute the asymptotic limit of the
error estimate $F$. To obtain this we fetch
from section \ref{s:treat} 
\begin{equation}
F=\sqrt{1-\frac{\int_0^{\xi_{BK}} b(\xi;\epsilon) d\xi }
{\int_0^{\xi_{BK}} b(\xi;0) d\xi}}  \label{eqn:Fbxi} \ \   ,
\end{equation}
where
\begin{equation}
\xi_{BK}=2R L_{BK} =A \; kR  \  \   ,
\end{equation}
cf eq. \EqRef{LBK}.
To begin with we are only interested in the leading asymptotic behavior of the function
$b(\xi;\epsilon)$.
From section \ref{s:treat} we therefore collect
\begin{equation}
b(\xi;\epsilon) \sim 
\frac{Z_2(z_0;\epsilon)}
{Z'(z_0;\epsilon)} e^{z_0 \xi}  \  \  , 
\end{equation}
valid for large values of $\xi$.
From section \ref{s:implBER} we find
\begin{equation}
z_0 \sim -\epsilon
\end{equation}
\begin{equation}
Z'(z_0;\epsilon) \sim 1
\end{equation}
\begin{equation}
Z_2(z_0;\epsilon) \sim Z_2(0,0)
\end{equation}
to leading order in $\epsilon$, 
for error bounds please go back to section \ref{s:implBER}.
Finally, from section \ref{s:penum}
we have
\begin{equation}
\epsilon =c  (kR)^{-2/3} \  \  .
\end{equation}

We are then in the position to compute the large $kR$ limit
of the error estimate, which we easily evaluate to
\begin{equation}
F\sim  \left( 1-\frac{1-e^{-\epsilon \xi_{BK}}}
{\epsilon \xi_{BK}}\right)^{1/2}  
= \left( 1-\frac{1-e^{-cA(kR)^{1/3}}}
{cA(kR)^{1/3}} \right)^{1/2}  \  \  . \label{eqn:final_F}
\end{equation}
We observe that this function will definitely approach
unity which implies that individual eigenstates ceases
to be resolved. This final collapse will occur on the scale
\begin{equation}
kR \sim (cA)^{-3}
\end{equation}
Note that if only one symmetry subspace is considered, then
$A\approx 1/8$,  so this might correspond to a very high energy.

But how is this asymptotic expression approached.

Actually we know that 
$F=0$ if $kR$ is less than some critical value which is given by
$\xi_{BK}=1/(1+\epsilon/2)$ (cf. eq. \EqRef{small_xi} ) or
written out explicitly
\begin{equation}
A\;kR=\frac{1}{1+\frac{c}{2}(kR)^{-2/3}}  \  \  .
\end{equation}
The solution is given by
\begin{equation}
(kR)_{\em threshold}=\left( \frac{1}{6}\alpha^{1/3}-c\; \alpha^{-1/3} \right)^3
\  \  ,
\end{equation}
where
\begin{equation}
\alpha=\frac{108}{A}+6\sqrt{6c^3+324/A^2}  \  \  .
\end{equation}
Unstable periodic orbit below this 
threshold are never diffractive.
Of course, neutral orbits in this range are subject to diffraction 
corrections but
we do only consider the error due to unstable orbits.
We
are thus unable to estimate the semiclassical
error for small values of $kR$.

\begin{figure}
\epsfxsize=11cm
\epsfbox{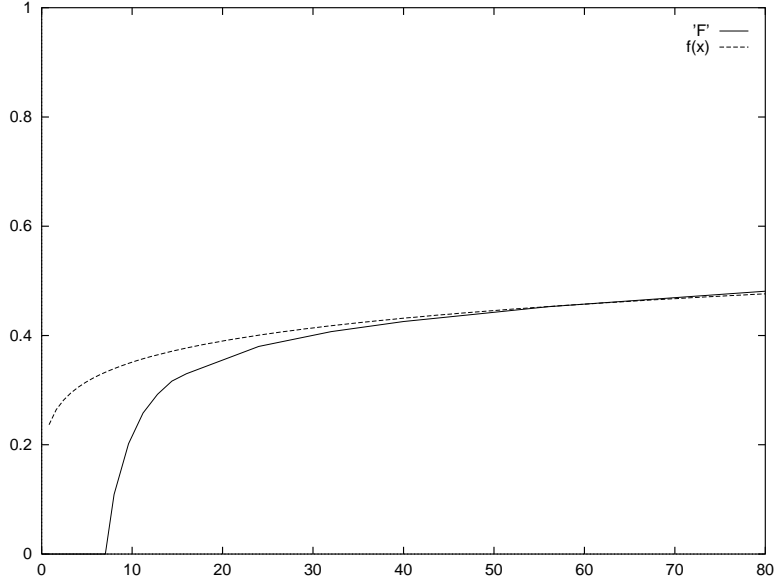}
\caption{The error estimate 
$F(kR)$ versus $kR$ according to eqs \EqRef{bapp} (full line)
compared with the asymptotic expression  \EqRef{Fbxi} (dashed line).
$A=1/8$ and $c$ is chosen to be $c=1$.}
\label{f:F}
\end{figure}

To estimate the intermediate behavior we introduce a further
approximation of eq. \EqRef{b}
\begin{equation}
b(\xi;\epsilon)= \frac{1}{2\pi i} 
\int_{\sigma-i\infty}^{\sigma+i\infty}\frac{Z_2(z;\epsilon)}{Z(z;\epsilon)}
e^{z\xi}dz
\approx Z_2(0,0)
\int_{\sigma-i\infty}^{\sigma+i\infty}\frac{1}{Z(z;\epsilon)}
e^{z\xi}dz  \label{eqn:bapp}
\end{equation}
which is highly reasonable, cf.\ eq.\ \EqRef{Z2_slow} . 
We replace the classical zeta function $Z(z,\epsilon)$
by the approximate one $\hat{Z}(z,\epsilon)$, the transforms are again computed by FFT 
technique and the resulting function $F(kR)$ is plotted in fig.
\ref{f:F}. In the computation we use $A=1/8$
and an arbitrary value of $c=1$.
We see is a very steep ascent at the threshold discussed above and
a fast approach to the asymptotic formula \EqRef{final_F}.
It is likely that the individual states ceases to be resolved
already here. However, it is hard to make
any safe prediction regarding the crudeness of our treatment
of the penumbra problem, see also the discussion of $c$ below.
However, the function  $F$ {\em does} approach unity, sooner or later,
and it is difficult to find an prevarication of this fact.

\vspace{0.5cm}

There is an issue which we appear to have forgotten, 
the constant $c$ in eq. \EqRef{cdef} is not really a constant,
it depends weakly on $kR$, $R$, $r_1$ and $r_2$,
see fig \ref{f:naka3} and appendix \ref{app:A}.
First, this dependence is to weak to be able to alter our general conclusions.
Moreover, it is not obvious how to implement the dependence on $r_1$ and $r_2$.
In the BER application we consider disk to disk segment whereas  in the
study of the circle Green function in sec \ref{s:penum}
we consider (square) boundary to boundary segments. 
To make the exact connection
one has to convolute the Circle Green function with itself, the outcome
of this operation is not obvious. 
So we used $c$ as if it was a constant when implementing the
BER approximation.
Therefore the "constant" $c$ 
coming out of the other end of the BER calculation
should have some residual weak dependence on $R$ and $kR$.

\section{Discussion of the validity of the various approximations involved}

\label{s:disc}

No chain is stronger that the weakest link.
The results presented here is based on a long series of approximations and 
assumptions. Some of them may be readily justified and should  hardly be controversial
but some may seem a lot more crude and one may ask if they will allow the chain to 
break.

\subsection{The correction to the semiclassical weights}
\label{assum:penum}

On one hand we argued that penumbra diffraction cannot be accounted for by
multiplicative corrections but on the other hand we {\em needed} multiplicative
corrections to be able to use the machinery of the periodic orbit theory
and evolution operators. So we simply constructed a multiplicative weight
inspired by the penumbra diffraction that should be able to provide us with an
estimate of the error in the Berry-Keating formula.
The procedure was discussed at some length in sec.\ \ref{s:penum}.

One could object that it is too crude to approximate the gradual
transition of the circle Green function 
with a step function, and suggest a more smooth weight.
This is in principle possible, but that would make the calculation in sec.\
\ref{s:implBER} immensely more complicated without changing the result
in any significant way.

\subsection{The BER approximation}
\label{assum:BER}

It is natural that a theory for the asymptotics of the periodic orbits
is hard to check numerically.
In ref. \cite{PDsin} we compared the 
exact trace formulas with those of the BER approximation for lengths up to 
roughly
the horizon
$1/2R$ for the Sinai billiard. The results were encouraging but hardly asymptotic.


However, a range of asymptotic predictions
of the BER approximation  appears to be correct. It does provide
the suggested exact diffusion behavior in the regular Lorentz gas 
(with
unbounded horizon) \cite{bleh92,PDlyap} and the related correlation decays \cite{PDRA}
as well as the small radius limit of the Lyapunov exponent\cite{PDsmall}.
We therefore feel confident that the approach works even if more rigorous results
are called for.

We introduced some extra approximations, valid in the small $R$ limit,
but our experience is that they work excellent even for rather large $R$.

One could again raise objection that the diffractive weight is discontinuous
and that this could cause problems for averages to settle down. But we saw in
section \ref{s:implBER} that the effect is just to change the pruning rules
slightly, the billiard is as discontinuous as before and the fluctuations of
the pseudo orbit sums appearing in the numerator and denominator \EqRef{Fb} 
are similar.
The integrals in  eq. \EqRef{Fb} are self averaging and the fluctuations of the
integrands irrelevant for the estimate of the error.

\subsection{The diagonal approximation}
\label{assum:diag}

The diagonal approximation underlying eq. \EqRef{D2} 
can be verified assuming that the spectral
statistics is given by Random Matrix Theories \cite{KeatPriv}. The diagonal approximation 
on the diffractive
sum behind \EqRef{dD2} is natural, in particular since the majority of pseudo orbits
are diffractive for small $\epsilon$,
but by no means obvious.

\vspace{0.7cm}

I am grateful to Jon Keating for a very useful discussion.
This work was supported by the Swedish Natural Science
Research Council (NFR) under contract no. F-AA/FU 06420-313.

\appendix

\section{Stationary phase analysis of the circle Green function}

\label{app:A}

We will consider scatterings in extreme forward angles so only the second
term in the integral \EqRef{Gmint} (m=0) contributes.
This means that we only need to evaluate the integral
\begin{equation}
G^{(0)}(r_1,r_2,\Delta\theta)=
\frac{i}{8}\int_{-\infty}^{\infty}S_\ell(kR) H_\ell^{(1)}(kr_1)
H_\ell^{(1)}(kr_2) e^{i\ell \Delta\theta}d\ell  \label{eqn:Jdef} \ \ ,
\end{equation}
where $0\ll \Delta\theta<\pi$. 
Actually, this integral is divergent but as long as we study the
stationary phase approximation, it serves our purposes.
We can still safely use the Debye approximation for the Hankel functions 
$H^+_\ell(kr_1)$ and
$H^+_\ell(kr_2)$  
\begin{equation}
H_\ell^{(1)}(z)\sim \sqrt{\frac{2}{\pi\sqrt{z^2-\ell^2}}}
e^{i[\sqrt{z^2-\ell^2}-\ell\; \arccos (\ell/z)-\pi/4]} \ \ ,
\end{equation}
because $kr_1 \gg \ell$ and $kr_2 \gg \ell$.
However,
the phase shift function needs a more careful analysis. 
The phase shift function $S_\ell (kR)$ is of unit modulus
and we call the phase $\gamma(kR,\ell )$
\begin{equation}
S_\ell(kR) = -\frac{H_\ell^{(2)}(kR)}{H_\ell^{(1)}(kR)}\equiv e^{i\gamma(kR,\ell)} \ \ .
\end{equation}
The Green function now reads
\begin{equation}
G^{(0)}(r_1,r_2,\Delta\theta)=\int_{-\infty}^{\infty}
A(r_1,r_2)
e^{i(\ell\Delta\theta - \Psi(\ell))}
d\ell  \ \ ,
\end{equation}
where the slowly varying amplitude
$A(r_1,r_2)$ is composed of the Hankel functions $H_\ell^{(1)} (kr_1)$
and $H_\ell^{(1)} (kr_2)$.
The asymptotics of the integral is
determined by the phase function
\begin{equation}
\Psi(\ell)=-\gamma(kR,\ell)-[\sqrt{(kr_1)^2-\ell^2}-\ell\; \arccos (\ell/(kr_1))]
-[\sqrt{(kr_2)^2-\ell^2}-\ell\; \arccos (\ell/(kr_2))]+\pi/2 \ \ .
\end{equation}

We will now investigate
the phase of $S_\ell(kR)$ in detail.
To this end we will use the uniform approximation for Hankel functions relating
the phase of Hankel functions 
to the phase of Airy functions according to \cite{AS}
\begin{equation}
S_\ell(kR)=-\frac{\mbox{Ai}(-x)+i\mbox{Bi}(-x)}{\mbox{Ai}(-x)-i\mbox{Bi}(-x)}
\equiv{e^{i\gamma(\ell,kR)}} \  \  ,
\end{equation}
where
\begin{equation}
x(kR,\ell)=\left\{ \begin{array}{cc}
[\frac{3}{2}(\sqrt{(kR)^2-\ell^2}-\ell \; \arccos(\ell/(kR)))]^{2/3} 
& \ell<kR\\
-[\frac{3}{2}(\ell\cdot\log (\frac{\ell+\sqrt{\ell^2-(kR)^2}}{kR})-
\sqrt{\ell^2-(kR)^2}]^{2/3} & \ell>kR \end{array}\right.  \ \ .
\end{equation}
Defining
\begin{equation}
\ell=kR(1+\epsilon)  \  \  ,
\end{equation}
we have for small $\epsilon$
\begin{equation}
-x=(\sqrt{2}kR)^{2/3} (\epsilon+O(\epsilon^2))  \label{eqn:x}
\  \  .
\end{equation}

Using this expression for $x$ one gets the so called transition region
approximation.
This approximation fails to yield the Debye approximation as its asymptotic limit
but the nice thing is that there is a considerable overlap 
because whenever $x\ll \ell^{2/3}$ we can use the transition region approximation and when
$x \gg 1$ we can use Debye. 

$\gamma(x)$ is
a complicated function but one has the following asymptotic expressions \cite{AS}
\begin{eqnarray}
\gamma(x) \sim  - e^{-\frac{4}{3}(-x)^{3/2}}(1+O(1/(-x)^{3/2})) &
x\rightarrow-\infty  \ \ , \label{eqn:minusinf} \\
\gamma(x) =-\frac{\pi}{2}-\frac{4}{3}x^{3/2}+O(x^{-3/2}) & x \rightarrow +\infty
\  \  ,
\end{eqnarray}
so we obtain the Debye approximation when $x \rightarrow \infty$
as expected.

\begin{figure}
\epsfxsize=8cm
\epsfbox{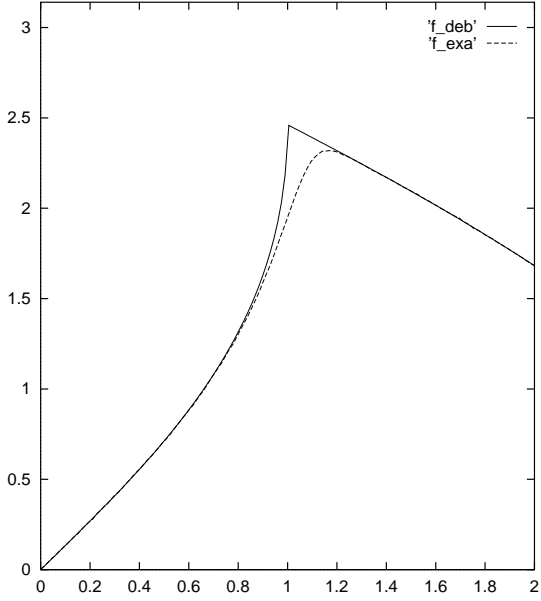}
\caption{The function $\Psi_\ell(\ell)$ versus $\ell$ for
fixed values of  $k=30$, $R=1$ and $r_1=r_2=3$ }
\label{f:Psil}
\end{figure}

The stationary phase condition will simply read
$\Psi_{\ell}(\ell)=\Delta\theta$ (subscripts denote
differentiation).
In fig.\ \ref{f:Psil} 
we plot the function $\Psi_\ell (\ell)$ versus $\ell$ for some arbitrary chosen
values of $r_1/R=r_2/R=3$ and $kR=30$ together with its Debye approximation.
The plot gives a picture of how the stationary phase
will perform for all $\Delta\theta$, just place the ruler 
horizontally at $\Delta\theta$. 
If it goes below the maximum, it intersects the curve twice.
The corresponding saddle points corresponds to the direct (larger $\ell$) and
reflected ray (smaller $\ell$) respectively.

We are interested in the location of the maximum of $\Psi_\ell (\ell)$.
This turns out the lie in the region where eqs \EqRef{x} {\em and} 
\EqRef{minusinf} applies.
To leading order in $(kR)$ the location of the maximum is the solution to the
equation
\begin{equation}
-xe^{-\frac{4}{3}(-x)^{3/2} }= 2^{-8/3} (kR)^{-1/3} 
\left( \frac{R}{\sqrt{r_1^2-R^2}} +\frac{R}{\sqrt{r_2^2-R^2}} \right)
\equiv (kR)^{-1/3} C(r_1,r_2)  \label{eqn:C} \  \  .
\end{equation}
By elementary methods one can show that the solution to this equation
lies in the range
\begin{equation}
\frac{9}{16} \left(\log \frac{(kR)^{1/3}}{C(r_1,r_2)}\right)^{2/3} < (-x)_{max} 
<\left(\log \frac{(kR)^{1/3}}{C(r_1,r_2)}\right)^{2/3}  \  \  .
\end{equation}
Where in this range the solution lies is completely irrelevant for us.
We use this liberty and
choose $\epsilon$ in \EqRef{dcrit} to be (cf. eq. \EqRef{x})
\begin{equation}
\epsilon_{max}=2^{-1/3} (kR)^{-2/3} 
\left(\log \frac{(kR)^{1/3}}{C(r_1,r_2)}\right)^{2/3}
\equiv c (kR)^{-2/3} \  \  .
\end{equation}
where
\begin{equation}
c= c(kR,R,r_1,r_2)=2^{-1/3} \left(
\frac{1}{3} \log kR +\frac{8}{3} \log 2 -
\log (\frac{R}{\sqrt{r_1^2-R^2}}+\frac{R}{\sqrt{r_2^2-R^2}})
\right)^{2/3}
\label{eqn:cdef}
\end{equation}

\newcommand{\PR}[1]{{Phys.\ Rep.}\/ {\bf #1}}
\newcommand{\PRL}[1]{{Phys.\ Rev.\ Lett.}\/ {\bf #1}}
\newcommand{\PRA}[1]{{Phys.\ Rev.\ A}\/ {\bf #1}}
\newcommand{\PRD}[1]{{Phys.\ Rev.\ D}\/ {\bf #1}}
\newcommand{\PRE}[1]{{Phys.\ Rev.\ E}\/ {\bf #1}}
\newcommand{\JPA}[1]{{J.\ Phys.\ A}\/ {\bf #1}}
\newcommand{\JPB}[1]{{J.\ Phys.\ B}\/ {\bf #1}}
\newcommand{\JCP}[1]{{J.\ Chem.\ Phys.}\/ {\bf #1}}
\newcommand{\JPC}[1]{{J.\ Phys.\ Chem.}\/ {\bf #1}}
\newcommand{\JMP}[1]{{J.\ Math.\ Phys.}\/ {\bf #1}}
\newcommand{\JSP}[1]{{J.\ Stat.\ Phys.}\/ {\bf #1}}
\newcommand{\AP}[1]{{Ann.\ Phys.}\/ {\bf #1}}
\newcommand{\PLB}[1]{{Phys.\ Lett.\ B}\/ {\bf #1}}
\newcommand{\PLA}[1]{{Phys.\ Lett.\ A}\/ {\bf #1}}
\newcommand{\PD}[1]{{Physica D}\/ {\bf #1}}
\newcommand{\NPB}[1]{{Nucl.\ Phys.\ B}\/ {\bf #1}}
\newcommand{\INCB}[1]{{Il Nuov.\ Cim.\ B}\/ {\bf #1}}
\newcommand{\JETP}[1]{{Sov.\ Phys.\ JETP}\/ {\bf #1}}
\newcommand{\JETPL}[1]{{JETP Lett.\ }\/ {\bf #1}}
\newcommand{\RMS}[1]{{Russ.\ Math.\ Surv.}\/ {\bf #1}}
\newcommand{\USSR}[1]{{Math.\ USSR.\ Sb.}\/ {\bf #1}}
\newcommand{\PST}[1]{{Phys.\ Scripta T}\/ {\bf #1}}
\newcommand{\CM}[1]{{Cont.\ Math.}\/ {\bf #1}}
\newcommand{\JMPA}[1]{{J.\ Math.\ Pure Appl.}\/ {\bf #1}}
\newcommand{\CMP}[1]{{Comm.\ Math.\ Phys.}\/ {\bf #1}}
\newcommand{\PRS}[1]{{Proc.\ R.\ Soc. Lond.\ A}\/ {\bf #1}}


\begin{thebibliography}{99}
%
{\small

\bibitem{Gut}  M.~C.~Gutzwiller, {\em Chaos in Classical and Quantum
                Mechanics}, Springer, New York (1990).
\bibitem{Heller} E.~J.~Heller, S.~Tomsovic and M.~A.~Sep\'{e}lveda,
        Chaos {\bf 2}, 105 (1992).
\bibitem{BK90}     M.~V.~Berry and J.~P.~Keating, \JPA{23}, 4839 (1990)
\bibitem{CE93}    P.~Cvitanovi\'{c} and B.~Eckhardt, \PRL{63}, 823 (1989).
                (1993).
\bibitem{Boas}    P.~A.~Boasman, Nonlinearity {\bf 7}, 485 (1994).
\bibitem{Bogo92} E.~B.~Bogomolny, Nonlinearity {\bf 5}, 805 (1992).
\bibitem{DahlRuss91} P.~Dahlqvist, \JPA{24}, 4763 (1991).
\bibitem{Freddy92} F.~Christiansen and P.~Cvitanovi\'{c}, Chaos {\bf 2}, 61 (1992).
\bibitem{Tanner91} G.~Tanner, P.~Scherer, E.~B.~Bogomolny, B.~Eckhardt and D.~Wintgen,
\PRL{67} 2410 (1991).
\bibitem{SS91} M.~Sieber and F.Steiner, \PRL{67}, 1941 (1991).
\bibitem{Prim98}   H.~Primack and U.~Smilansky,  \JPA{31}, 6253 (1998).
\bibitem{Saraceno} M.~Saraceno and A.~Voros, Chaos {\bf 2}, 99 (1992).
\bibitem{ghost}   B.~Sundaram and R.~Scharf, \PD{83}, 257 (1995).
\bibitem{Sieb97}  M.~Sieber, N.~Pavloff and C.~Schmit,
                   \PRE{55}, 2279 (1997).
\bibitem{Sieb97a}  H.~Schomerus and M.~Sieber, \JPA{30}, 4537 (1997).
\bibitem{Tanner} G.~Tanner, \JPA{30}, 2863 (1997).
\bibitem{Prim95}  H.~Primack, H-~Schanz, U.~Smilansky and I.~Ussishkin,
                  \PRL{76}, 1615 (1996).
\bibitem{BER}   V.~Baladi, J.~P.~Eckmann and D.~Ruelle,
                Nonlinearity {\bf 2} (1989) 119.
\bibitem{PDreson}P.~Dahlqvist,  \JPA{27}, 763 (1994).
\bibitem{PDsin} P.~Dahlqvist, Nonlinearity {\bf 8},11 (1995).
\bibitem{PDzak} P.~Dahlqvist, J. Techn. Phys. {\bf 38}, 189 (1997).
\bibitem{PDlyap} P.~Dahlqvist, \JSP{84},  773 (1996).
\bibitem{CAT1}   J.~H.~Hannay and M.~V.~Berry, \PD{1}, 267 (1980).
\bibitem{CAT2}   J.~P.~Keating, Nonlinearity {\bf 4}, 309 (1991).
\bibitem{geo}    M.~C.~Gutzwiller, \PRL{45}, 150 (1980).
\bibitem{PDnaka}  P.~Dahlqvist, Chaos Solitons and Fractals {\bf 8}, 1011, (1997).
\bibitem{DasBuch} P.~Cvitanovi\'{c} et.al. 
{\em Classical and Quantum Chaos: A Cyclist Treatise},
                  http://www.nbi.dk/ChaosBook/, 
                  Niels Bohr Institute (Copenhagen 1997).
\bibitem{prob} C.~F.~F.~Karney, \PD{8}, 360 (1983).
\bibitem{BIM}     M.~V.~Berry and M.~Wilkinson, \PRS{A392}, 15 (1984).
\bibitem{onedisk} B.~R.Levy and J.~B.~Keller, Comm.\ Pure.\ Appl.\ Math,
{\bf 12}, 159 (1959).
\bibitem{Kai}    K.~T.~Hansen, {\em Symbolic Dynamics in Chaotic
                 Systems}, Ph.D. thesis, Oslo (1900).
\bibitem{AAC90}   R.~Artuso, E.~Aurell and P.~Cvitanovi\'{c},
                Nonlinearity {\bf 3}, 325 and 361, (1990).
\bibitem{PDsmall}P.~Dahlqvist,  
                  Nonlinearity {\bf 10}, 159 (1997).   
\bibitem{bleh92}  P.~M.~Bleher, \JSP{66}, 315, (1992).
\bibitem{PDRA}     P.~Dahlqvist, R.~Artuso, \PLA{219}, 212 (1996).
\bibitem{KeatPriv} J.~P.~Keating, private communication.
\bibitem{AS}  M.~Abramovitz and I.~A.~Stegun, {\em Handbook of
                mathematical functions}, Washington: National
                Bureau of Standards, (1964).
} 
\end{thebibliography}
\end{document}